\documentclass[useAMS,usenatbib]{mn2e}
\usepackage{psfig}

\def\etal{{\rm et~al.\ }}
\def\hmpc{\;h^{-1}{\rm Mpc}}

\def\hkpc{h^{-1}{\rm kpc}}
\def\kms{{\rm \;km\;s^{-1}}}

\def\kmsmpc{\kms\;{\rm Mpc}^{-1}}
\def\msun{{\rm h^{-1} M_{\odot}}}
\def\msolar{M_{\odot}}

\def\simlt{\lower.5ex\hbox{$\; \buildrel < \over \sim \;$}}
\def\simgt{\lower.5ex\hbox{$\; \buildrel > \over \sim \;$}}

\title[Radiation-Induced Large-Scale Structure]{
Radiation-induced large-scale structure during
the reionization epoch: the autocorrelation function
}

\author[R.A.C. Croft and G. Altay]{
Rupert A.C. Croft$^{1}$\thanks{E-mail: rcroft@cmu.edu} and
Gabriel Altay$^{1}$\\
Dept.   of  Physics,   Carnegie   Mellon  University,
Pittsburgh, PA 15213, USA\\ 
}
\begin{document}
\pagerange{\pageref{firstpage}--\pageref{lastpage}} \pubyear{2005}

\maketitle

\label{firstpage}

\begin{abstract}
The structures produced during the epoch of reionization by the
action of radiation on neutral hydrogen are in principle different
to those that arise through gravitational growth of initially
small perturbations. We explore the difference between the
two mechanisms using high resolution cosmological radiative transfer.
Our computations use a Monte Carlo code which raytraces directly
through SPH kernels without a grid, preserving the high
spatial resolution
of the underlying hydrodynamic simulation.
Because the properties of the first sources of radiation are  uncertain,
we simulate a range of models with different source properties
and recombination physics. We examine the morphology of 
the neutral hydrogren distribution and the reionization history in these
models. We find that at fixed mean neutral fraction,
structures are visually most affected by the existence of a lower
limit in source luminosity, then by galaxy mass to light ratio,
and are minimally affected by changes in 
the recombination rate and amplitude of mass fluctuations.
We concentrate on the autocorrelation function of the neutral
hydrogen, $\xi_{HI}(r)$ as a basic
quantitive measure of Radiation Induced Structure (RIS.)
 All the models we test exhibit
a characteristic behaviour, with $\xi_{HI}$ becoming 
initially linearly antibiased with respect to the matter correlation
function, reaching a minimum bias factor $b\sim0.5$ when the universe
is $\sim 10-20\%$ ionized. After this  $\xi_{HI}$ increases
rapidly in amplitude, overtaking the matter correlation function.
It keeps a power law shape,
but flattens considerably, reaching an
asymptotic logarithmic slope of $\gamma\simeq-0.5$. 
The growth rate of $HI$ fluctuations is exponentially more rapid
than gravitational growth over a brief interval of redshift 
$\Delta z \sim 2-3$.

\end{abstract}
 
\begin{keywords}
Cosmology: observations -- large-scale structure of Universe
\end{keywords}

\section{Introduction}

In the standard cosmological model, the large-scale structure in the
density field grows from small initial perturbations
through the mechanism of 
gravitational instability.
Statistical measures of this structure can be used to 
both verify the growth mechanism (see e.g.,
Bernardeau \etal 2002 and references therein) and quantify the 
initial pertubations. A different kind of growth of 
structure is expected when we consider the neutral hydrogen density field
during the epoch of reionization (see e.g., the review by Loeb \& 
Barkana 2001). In this case, bubbles of
ionized material form first around bright sources and grow as the ionization
fronts overlap until the universe is fully ionized.
Statistical measures applied to this ``Radiation Induced Structure''
(hereafter RIS) can be used in a similar way to the gravitational
instability picture above, but this time to verify the process of 
reionization and the nature of the sources of radiation. The effect
of RIS is likely to be qualitatively different from that of
gravity, and as a result the statistical signatures will be different. 
Our aim in this paper is to explore the differences, using ray traced
simulations of reionization.  We aim to both 
find out how reionization is different from gravity in the way
it forms structure
and how to use these differences to categorize reionization scenarios.
The statistical properties of RIS  have been explored in many other
works, e.g., using the power spectrum of HI fluctuations by 
Furlanetto \etal (2004), Zaldarriaga \etal (2004), Morales \& Hewitt (2004),
and using Minkowski functionals by Gleser \etal (2006).
In the present paper we will focus on the autocorrelation function.

Theoretical studies of reionization include both
analytic work, e.g., Miralda-Escud\'{e} \etal (2000), 
Wyithe \& Loeb (2003),
Cen (2003),
Liu \etal (2004),
Furlanetto \etal (2004),
 and numerical simulations, e.g., 
Razoumov \& Scott (1999),
Abel \etal (1999),
Gnedin (2000),
Ciardi \etal (2001),
Sokasian \etal (2001),
Razoumov \etal (2002), 
Sokasian \etal (2004).

Recently, N-body simulations with radiative transfer post
 processing have been performed by 
Iliev \etal (2006), and
Zahn \etal (2007) 
with box sizes as large as
$100 \hmpc$ and able to resolve halos down to  masses of
$\sim 2 \times 10^9 \msolar$. 
Kohler \etal (2007) have performed simulations with an extremely
large box size
(up to $1280 \hmpc$) in which 
hydrodynamics and radiative transfer are coupled self consistently.
These simulation rely on 
higher resolution simulations to calibrate the sub grid  ($< 10 \hmpc$)
physics. 
Other recent advances in simulating reionization include the work of 
Trac \& Cen (2006), a hybrid N-body dark matter and RT approach,
 a code which was used to study the growth of bubbles during reionization
by Shin \etal (2007).

Analytic work has been carried out using perturbation
theory to predict inhomogeneities in the density of neutral hydrogen
and photons (Zhang \etal 2007), as well as Press-Schechter based
analyses (e.g., Furlanetto \etal 2004.) Most relevant to the work here,
the autocorrelation function of 21cm emission has been examined
in analytic models by Wyithe \& Morales (2007) and Barkana (2007).

In the present paper we do not aim to simulate particular observational
probes of this epoch, such as the Doppler scattering of CMB photons on
relativistic electrons, or the 21cm emission from neutral hydrogen.
Instead, we concentrate on the differences between $\xi(r)$ for the
density field and the density field modulated by  RIS. In principle, this
will be directly observable
in the future, through the various observational probes
(e.g., Carilli \etal 2004, Peterson \etal 2005, Vald\'{e}s \etal 2006)
 We
leave exploration of this to future work.

The format of this paper is as follows. In \S 2 we describe
the simulations, including the N-body outputs and the radiative transfer
code. We also describe our 
different models for the sources of ionizing radiation and 
different physical conditions we simulate. In \S 3 we show how global
properties such as the mean ionized fraction evolve as a function
of redshift in the different runs, and then in \S 4 examine the
morphology of the neutral and ionized hydrogen density field.
In \S 5 we measure the autocorrelation function in our different models and
explore how it can differentiate between them. We summarise our results
and discuss them in \S 6

\section{Simulations}

We work in the context of the standard cosmological constant dominated 
universe,
 with parameters $\Omega_{\Lambda}=0.7$, $\Omega_{\rm m}=0.3$
$\Omega_{\rm b}=0.04$, and a Hubble constant $H_{0}=70 \kmsmpc$. The
initial linear power spectrum is cluster-normalised with a linearly
extrapolated amplitude of $\sigma_{8}=0.9$ at $z=0$. The radiative transfer
is carried out as post-processing on N-body simulation outputs. 

\subsection{N-body outputs}

We run our N-body simulations with the cosmological hydrodynamic code
Gadget (Springel \etal 2001). Our fiducial simulation is run 
in a $40 \hmpc$ cubical volume.
 Iliev \etal (2006) have shown in tests using subvolumes
of a larger (100 $\hmpc$) simulation that $30 \hmpc$ is the smallest box
side length for which the scatter in reionization histories in different
volumes is reasonably small.
We use $256^{3}$ dark matter and $256^{3}$
gas particles. The mass resolution is therefore $6.05 \times 10^7 \msolar$ 
for gas particles and $3.93 \times 10^8 \msolar$ for dark matter. We also 
run higher resolution models in tests, as detailed below.
 We do not include radiative cooling or star formation when
computing the gas dynamics so that our simulations
are similar to those carried out by 
Sokasian \etal (2001). We run the RT as postprocessing, so that there is
no coupling between the hydrodynamics and radiation. In this
respect, the gas serves to trace the dark matter distribution closely (there
is little difference at these high redshifts.)
We chose our sources of radiation to be associated with dark matter haloes
(see below).

In addition to the fiducial run, we run another model with identical box size
and particle number but with different random initial phases in order
to roughly indicate the effect of simulation cosmic variance.
We also run a model with a different amplitude of mass
fluctuations ( $\sigma_{8}=0.7$.)
 For resolution
tests, we run a simulation with $128^{3}$ gas and dark matter
particles in a box of size $20 \hmpc$ ( the same mass resolution as the fiducial run)
and a simulation with 
$256^{3}$ gas and dark matter particles in a box of size $20 \hmpc$ 
( eight times better mass resolution than the fiducial run).
All models were started at $z=50$ and run until $z=5.2$. We output 
snapshots of the density field every 25 Myr, so that there are approximately
40 snapshot files per run.

\subsection{Radiative transfer}

After choosing models for the sources of radiation (see below) we carry out
raytracing simulations of radiative transfer (RT) to study the evolution of
the neutral hydrogen density. The code we use to do this carries out
Monte Carlo RT to follow photon packets through the
distribution of matter. The code is based on that used by Croft (2004)
to study the fluctuating radiation background field at lower redshift but 
incorporates time dependent RT in a Monte Carlo manner similar to 
the CRASH code of Maselli \etal (2003) (we actually only treat
hydrogen here, so we are in fact closer to the earlier
work of Ciardi \etal (2001).
The code in the present
paper traces directly through the SPH particle kernels (see also 
Kessel-Deynet \& Burkert, 2000,
Susa, 2006,
Yoshida \etal 2007, and
Daqle \etal 2007, for non grid based radiative transfer, and
Semelin \etal 2007, and 
Oxley \& Woolfson 2003 for SPH codes that trace through a Barnes-Hut tree)
and so 
requires no regridding of the density field between outputs. The spatial
resolution of the RT is therefore in principle higher in dense regions
than would be possible with a uniform grid. This approach is
described in detail in Altay \etal (2007), where test
problems are carried out. We also outline 
some of the relevant features of the code briefly below.

In the present paper we model only the hydrogen component of the
Universe (assumed to comprise 0.76 of the baryonic mass). We also do not
explicitly follow the temperature evolution of the gas, beyond taking
temperatures of particles to be $10^{4}$ K when they are ionized and
$T_{\rm CMB}$ when they are not. We follow collisional ionization,
photoionization and recombinations using the rates given in 
Cen (1992).
We randomly 
sample source photons from a power law distribution, $F_{\nu}=\nu^{\alpha}$,
of photon energies (more details on the sources are given in 
section 2.3, below.) In the present work, photon packets are monochromatic
and are emitted isotropically from sources, again using a random
number generator to pick directions.

Every time a ray is traced through a particle,
the number of recombination photons which have been produced 
in that particle since the last time it was visited are added to a stack.
When the stack size reaches one packet, a recombination photon packet is
emitted from the particle where this occurs. The frequency of the
recombination radiation is given by the Milne relation 
(Osterbrock, 1989.)

We have one numerical code
parameter, $c_{f}$ which sets the size of the photon packet,
and therefore the time resolution of the code. This parameter $c_{f}$
is the number of fully neutral
simulation gas particles which could be ionized by one
photon packet if the energy in that packet was split up into 
13.6 eV photons. For example, if a packet consists of $N_{\gamma}$
13.6 eV photons and a gas particle contains $N_{HI}$ neutral
hydrogen atoms then $c_{f}=N_{\gamma}/N_{HI}$.  By trying runs 
with different values
of $c_{f}$ we change how well the code can resolve the recombination
timescale by governing the average interval between rays visiting 
particles and updating their ionization states. Packets of recombination
radiation as well as source radiation are governed by  $c_{f}$ so that 
shot noise arising from the discreteness of recombination 
modelling can be controlled.
We choose a suitable value of $c_{f}$  by carrying out convergence
tests (see e.g., \S 5.1). In practice we find that  $c_{f}=0.33$ is 
adequate for modelling clustering of HI and is the value we use in 
our fiducial simulations. This results in from $0.5-1.0 \times 10^{8}$
photon packets being used in each of the runs.

The paper Altay \etal (2007) is mainly concerned with presenting 
a closely related code, the publically available code {\small SPHRAY},
which has many additional features to the one used in this paper,
including the ability to model temperature evolution and Helium species.

\subsection{Sources/runs}

Eventually, observations of the RIS at the epoch of reionization
will be useful as probes of the sources of radiation as well as cosmology.
The statistical measures of this structure are likely to 
be correlated with the properties of the sources, their luminosities,
lifetimes, and clustering. One of the goals of this paper is to study
the RIS produced by various extremely simple models for the ionizing
source population, in order to see how they can be differentiated
(principally through the autocorrelation function, which we
focus on) and which features appear to be generic to the models
considered. 

The range of possible sources for reionization is extremely wide, 
including  decaying dark matter (e.g., Mapelli \etal 2006),
 primordial black holes (Ping \& Fang 2002, Ricotti \etal 2007),
high redshift miniquasars (e.g., Madau \etal 2004),
 population III stars (e.g., Sokasian \etal 2004),
 population II stars (e.g., Sokasian \etal 2003), some more
or less likely than others. Rather than attempting to simulate
particular models in detail, we restrict ourselves to simply parametrized
 models which relate
the ionizing radiation intensity directly to the dark matter distribution.
This is on the understanding that in most reasonable models of reionization
there would be some relationship between the two (either
through galaxies and stars associated with dark matter overdensities,
or directly through dark matter clumps decaying to ionizing photons).
In particular, we associate sources of radiation to dark matter halos.

This approach has been used also by Mellema \etal (2006),  who
use a constant mass to light ratio to assign ionizing radiation to 
each halo, as well by Zahn \etal (2007), who populate each 
halo with a single ionizing source whose luminosity is proportional
to host halo mass. McQuinn \etal (2006) have also
simulated 17 different variations
of this type of model
with various ionizing photon efficiencies, prescriptions for feedback and
and minihalos. McQuinn \etal (2006) focus on the morphology of HII regions.

All the runs we use for the main studies in this paper have a
simulation box length of $40 \hmpc$ and particle number of $2\times256^{3}$,
although as explained in \S 2.1, for resolution studies we have
some runs with different mass resolutions and box sizes.
We find halos using a standard friends-of-friends routine, 
with a linking length of 0.2 times the mean interparticle
separation. The minimum halo mass we use as a source 
in our fiducial run is $1.6\times10^{9} \msun$, containing only 8 (gas and
dark matter) particles.
 This is approaching the scale 
of mini halos expected to host numerous weak ionizing sources and provide small
scale clumpiness to the IGM, but as with the
calculations of Zahn \etal (2007) (who do have a smaller particle
mass) it is still approximately an 
order of magnitude too large.  
Although we only resolve such halos with a small number 
of particles, we have checked using
simulations with 8 times better mass resolution (see \S 5.1) that this 
does not affect our computation of the autocorrelation function of 
neutral hydrogen, the statistic we focus on. In general, the limited mass
resolution of simulations will affect results both through the absence
of sinks (minihalos) and sources hosted by small halos. Recent simulations
(e.g., Santos \etal 2007) are beginning to address this directly through
increases in simulation particle number. We return to this point in \S5.1 
and \S6.2.

We use the same density field for 10 of the 12 main runs,
carrying out the RT as postprocessing using different
source prescriptions. The other two are a low
fluctuation amplitude model ($\sigma_{8}=0.7$) run with the same
random phases and another model with the same amplitude as the 
fiducial case but with different phases. The 12  runs are differentiated by
their different halo mass to light ratios, treatment of the relationship 
between halo mass and ionizing radiation, the recombination rate, and spectrum
of radiation. An overview is given in Table \ref{runs}, and they are
descibed in detail below. We label the different simulations
by short descriptive names rather than numbers or letters in order
to avoid the necessity of the reader referring back to a table when
examining the results.

\begin{table*}
\caption[runs]{\label{runs}
Radiative transfer simulations
 discussed in this paper. Further details are given
in section 2.3}
\begin{tabular}{ccccccc}
\hline&\\
 Simulation & (Luminosity & Spectrum & Recomb- & Box length & Particle&
  Comments \\
  & integrated to   & $\alpha$& -ination rate & $(\hmpc)$ & number  &\\
  & $z=6$) /fiducial & $(F_{\nu}\propto\nu^{\alpha}$) &   &   &   &\\
\hline & \\
fiducial & 1.0 & $-4.0$ & 1.0 & 40.0 & $2\times256^{3}$ & \\
L/2 & 0.5 & $-4.0$ & 1.0 & 40.0 & $2\times256^{3}$ & \\
L/4 & 0.25 & $-4.0$ & 1.0 & 40.0 & $2\times256^{3}$ & \\
L/8 & 0.125 & $-4.0$ & 1.0 & 40.0 & $2\times256^{3}$ & \\
L indep M & 1.0 & $-4.0$ & 1.0 & 40.0 & $2\times256^{3}$ & luminosity
independent
 of halo mass\\
$M > 10^{10} \msun$ & 1.0 & $-4.0$ & 1.0 & 40.0 & $2\times256^{3}$ & 
lower limit on halo mass for sources\\
$M < 10^{10} \msun$ & 1.0 & $-4.0$ & 1.0 & 40.0 & $2\times256^{3}$ & 
upper limit on halo mass for sources\\
$\sigma_{8}=0.7$ & 0.87 & $-4.0$ & 1.0 & 40.0 & $2\times256^{3}$ & 
low amplitude of mass fluctuations\\
no recomb. & 1.0 & $-4.0$ & 0.0 & 40.0 & $2\times256^{3}$ & \\
2 $\times$ recomb & 1.0 & $-4.0$ & 2.0 & 40.0 & $2\times256^{3}$ & \\
$\nu^{-2}$ spectrum & 1.0 & $-2.0$ & 1.0 & 40.0 & $2\times256^{3}$ & \\
other ran. phases & 1.0 & $-4.0$ & 1.0 & 40.0 & $2\times256^{3}$ & 
same params. as fiducial, different ICs\\
hires & 1.0 & $-4.0$ & 1.0 & 20.0 & $2\times256^{3}$ & \\
small & 1.0 & $-4.0$ & 1.0 & 20.0 & $2\times128^{3}$ & \\
different $c_{f}$ & 1.0 & $-4.0$ & 1.0 & 40.0 & $2\times256^{3}$ & 
3 runs: photon packet $c_{f}=3.0, 1.0, 0.1$ \\
\hline&\\
\end{tabular}

\end{table*}

\indent $\bullet$ The {\bf fiducial} run.
Here the instantaneous luminosity 
of sources is proportional to the dark matter halo mass,
with $L=L_{0} \times M_{\rm halo}/\msun$
 In our fiducial run, we take $L_{0}=2.7\times10^{31}$erg/s/$\msun$,
and a  source spectrum with $F_{\nu}\propto\nu^{-4}$,
appropriate for  population II stellar sources (e.g., Sokasian \etal 2003).
This simple model is also used by  Zahn \etal (2007), who assume 
a number of photons proportional to 
halo mass, (with a conversion factor $3.1\times10^{41}$ photons/sec/$\msun$.)
This corresponds to our fiducial model having
a luminosity per halo mass 4 times larger than that 
of Zahn \etal. 
Using the computation of Zahn \etal, the source output of our fiducial
 model can be considered to be very roughly equivalent to one with Population
II stars forming with an efficiency of $f_{*}=0.1$
from a Salpeter IMF, with a stellar lifetime of $\Delta t= 5 \times 10^{5}$
 yrs and an escape fraction of $f_{\rm esc}=0.04$. 
 The recombination rate in the fiducial simulation is taken to
be that computed directly from the gas density, and 
diffuse recombination photons are treated. 

\indent $\bullet$ Runs {\bf L/2, L/4} and {\bf L/8}. These are the
same as the fiducial run in every respect except that the 
ionizing luminosity has been reduced by an overall factor of 2,4 or 8. These
models can be considered to be equivalent to for example reducing the
efficiency of star formation, and/or the ionizing escape fraction
 with respect to the fiducial run. The L/4 model therefore corresponds
to the model simulated by Zahn \etal (2007).

\indent$\bullet$ {\bf L indep M} We use the same halo list as the
fiducial run, but instead of a luminosity proportional
to halo mass, we assign the same luminosity to all halos. The total
integrated luminosity to $z=6$ is set to be equal to that in 
the fiducial run.

\indent$\bullet$ {\bf M $>$ 10$^{10}$ $\msun$, M $<$ 10$^{10}$ $\msun$}
For these two runs, the same
halo source list is used as in the fiducial run, but with either
an upper or lower cutoff applied in the mass of a halo which can host
a source. As in the previous model above, the total
integrated luminosity to $z=6$ is set to be equal to the fiducial run.
These runs can be considered to roughly model the effects of feedback which 
might cause disruption of galaxy sources in small halos.

\indent$\bullet$ {\bf $\sigma_{8}=0.7$}
This run is the same as the fiducial run, except using a simulation
which has a significantly lower amplitude of mass fluctuations, leading
to later halo formation times. The source luminosity was proportional
to the halo mass in the same fashion as for the fiducial run, also with 
 $L_{0}=2.7\times10^{31}$erg/s/$\msun$. The total integrated luminosity to
$z=6$ is therefore less than in the fiducial run.

\indent$\bullet$ {\bf No recomb, 2$\times$ recomb}
In these runs, the recombination rate of ionzed hydrogen
was either set to zero or doubled, and the number of recombination
photon packets adjusted accordingly. Otherwise, the runs are the same
as the fiducial run. These models can be thought of as parametrizing
the effects of changing the clumping factor of unresolved gas. 

\indent$\bullet$ {\bf $\nu^{-2}$ spectrum} 
A harder spectrum than the $\nu^{-4}$ spectrum used for the fiducial
run was used here. The total number of ionizing 
photons integrated to $z=6$ was set to be the same as the
fiducial run, so that $L_{0}=1.0\times10^{32}$erg/s/$\msun$.
This run gives a rough indication of the effects of a harder
spectrum than that of Population II stars, although it was not chosen
to reproduce the spectrum of any particular source. For example,
 at the ZAMS, Pop III stars are expected to have a spectrum with  
 $F_{\nu}\sim\nu^{\alpha}$, with
$\alpha\simeq -1.3$
close to the hydrogen ionizing edge (Tumlinson \etal 2003).
For representative composite spectra of AGN (e.g., Telfer \etal 2002) 
a value of  $\alpha=-1.8$ is possible.
The effect of the harder spectrum will be to allow the photons
to penetrate further. The ``preheating'' effect of hard photons is not
treated here as we do not follow the temperature evolution of the gas
explicitly.

\indent$\bullet$ {\bf Other random phases}
As a rough indicator of the effects of cosmic variance, this run has the
same parameters as the fiducial run, but uses a different
underlying density field realization.

We evolve all the simulations to redshift $z=5.5$, irrespective of
whether they have achieved full reionization by then.

\section{Global evolution}

Because we have many different simulation runs with different treatments
of recombination and models for the sources, we need to decide at which 
redshifts to compare our measures of clustering. The different runs
reach $50 \%$ mean ionized fraction by mass, $x_{m}$ at 
redshifts ranging from $z=9-6$, so that comparing them at the same
redshift will mean comparing different stages of reionization. We will
therefore concentrate in the paper on snapshots taken at the
same mean ionized fractions in the different runs, rather than at the
same redshifts.

\begin{figure}
\centerline{
\psfig{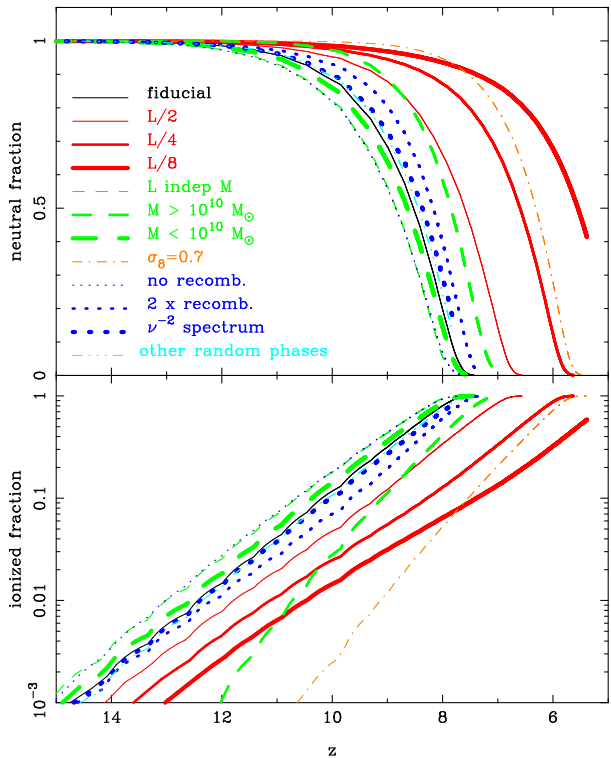}
}
\caption{Evolution of the mean mass weighted hydrogen neutral fraction (top
panel) and ionized fraction (bottom panel) with redshift.
Note that the bottom panel is on a log scale.
We show results for the 12 different models desribed in \S2.3.
\label{nvsz}
}
\end{figure}

In Figure \ref{nvsz} we show how the mean mass weighted neutral
fraction (linear scale) and mean mass weighted ionized fraction
(log scale) vary with redshift for the different runs. As expected,
 the run without recombinations reionizes first, reaching a $1\%$
ionized fraction by redshift $z=13$ and being $1\%$ neutral
at $z=7.8$. As reionization proceeds, the relationship between
mean ionized fraction $x_{m}$ and redshift is approximately exponential,
$x_{m}=e^{-(z-z_{i})}$, (here $z_{i}$ is the 
redshift when the model is fully ionized) for this and the other models. 
This is roughly true for all models, except for the model
where only large galaxies (halo masses $M > 10^{10} \msun$) are
sources, and the model with $\sigma_{8}=0.7$, both of
which have a substantially more rapid change in ionization with redshift,
$x_{m}=e^{-1.6(z-z_{i})}$. This can
be explained by the fact that galaxies massive enough to 
be sources only form relatively late  in these two models.

The models where the luminosity per dark matter halo atom was varied,
 fiducial, $L/2$, $L/4$, $L/8$ reach the  $x_{m}=0.5$ point later by 
$\Delta z=1$ for each halving of the luminosity. The slope of $\log
x_{m}$ vs $z$ is very similar for $x_{m}<0.1$ for these models but then 
changes as reionization proceeds. We note that the $L/4$ model reaches
$x_{m}=0.5$ at $z \sim 7$, similar to the model of Zahn \etal (2007),
which has a similar spectrum and source luminosity (see \S 2.3 above.)

\begin{figure}
\centerline{
\psfig{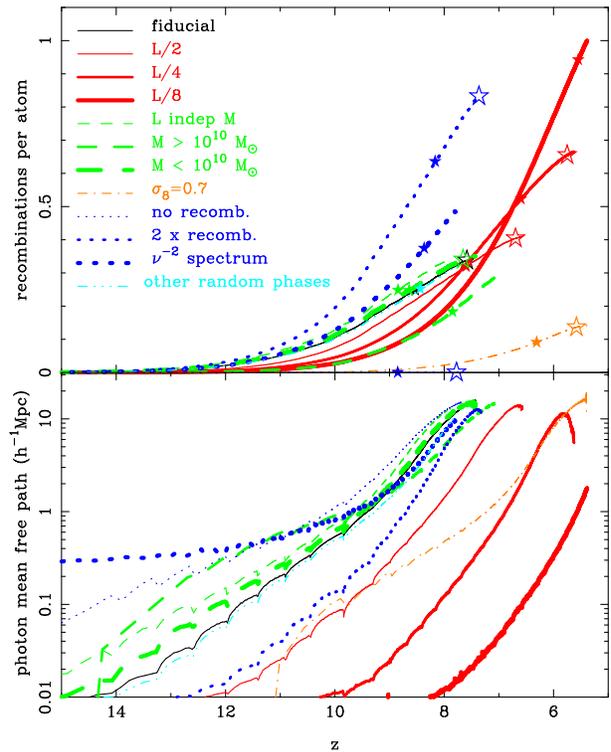}
}
\caption{Evolution of number of recombinations
per atom (top panel) and the photon mean free path
(bottom panel) with redshift.
In the top panel, we use the filled stars to mark the points when 
the mean mass weighted ionized fraction $x_{m}=0.5$
and the open stars to mark $x_{m}=0.99$ (not all runs
have been carried out to this point.) We show results for the 12 simulation
runs described in \S 2.3.
\label{recpath}
}
\end{figure}

More physical insight can be gained by looking at two other quantities as
a function of redshift, the number of recombinations per atom and the
photon mean free path, which are plotted in Figure \ref{recpath}.
To compute the former, we divide the cumulative number of recombination
photons by the total initial number of hydrogen atoms in the simulation
volume.
We plot a symbol on the curves at the point when the ionized fraction
by mass reaches $x_{m}=0.5$ and another at $x_{m}=0.99$ (not
all models reach this stage).
 Most of the models have below 0.5 recombination photons per atom
by the time they have fully reionized, indicating that recombinations
do not play the major role in the process of reionization. 
The curves for the different models tend to flatten off slightly 
for the last half of the reionization process. One striking feature
of these curves is the importance of the amplitude of mass
fluctuations, $\sigma_{8}$. As we would expect, the low amplitude model 
($\sigma_{8}=0.7$), having less clumping has a lower recombination rate.
When compared at the point when $x_{m}=0.5$, this model 
has experienced $3.2$ times fewer recombinations than the fiducial model.
This is greater than the ratio of $\sigma_{8}^{2}$ for the two models (1.65),
due to nonlinearity of clustering and the greater cosmic time
in the low amplitude model to get to this point.

The photon mean free path as a function of redshift is also shown in Figure 
\ref{recpath}. The mean path length between absorptions has been calculated by
averaging over the most recent $5\times10^{4}$ photon packets,
a relatively small number. As a result, sawtooth features in the
curves can be seen, which correspond to the times at which the
new density field snapshots were inputted (every 25 Myr). We have tested with
more widely spaced outputs (50 Myr) that this does not affect the underlying
curves.  Another artifact of the measurement is the behaviour at large mean
free path. When averaging the path length, we only include photon packets
that have not wrapped more than once round the box. As a result, when the
volume is close to fully ionized, the average is taken over a biased
sample, those photons
which have run into an absorber in less than a box length.  The mean 
free path we plot therefore falls. 

The overall behaviour of the mfp is as we would expect, with a rapid
rise for most models at early times, with a mfp corresponding
roughly to the size of the HII regions around sources. For example
at redshift $z=14$, when the ionized fraction is $\sim 10^{-3}$,
the mfp in the fiducial model is $\sim 10 \hkpc$. In the $\nu^{-2}$
spectrum case, the low cross section of neutral hydrogen for
the hard photons means that the mfp starts off nearly an
order of magnitude larger, $\sim 300 \hkpc$. The hardest photons will
travel much farther than this mean, and there is a low level of ionization
(ionized fraction $x_{m} \simgt 10^{-4} $)
present throughout  the whole volume, even at these very early redshifts.

\begin{figure}
\centerline{
\psfig{file=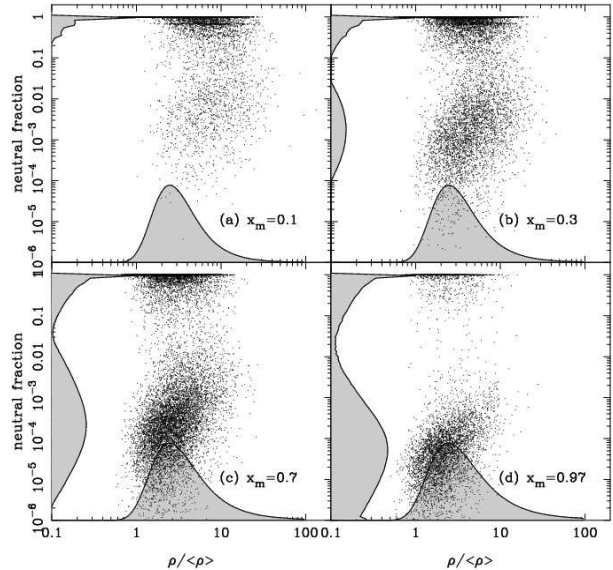,angle=-90.,width=8.0truecm}
}
\caption{
Scatter plot of gas density (in units of the mean) against 
hydrogen neutral
fraction for particles in the fiducial simulation run (see \S 2.3). We
show results for 4 different output times, characterized by
the mean mass weighted ionized fraction $x_{m}$ which appears in the
panel labels. In each panel, we also show as shaded areas
 histograms of the number of particles
in bins of hydrogen neutral fraction and also histograms binned by
density, $\rho/<\rho>$. The height of each histogram bin is on a log
scale.
\label{scatterfid}
}
\end{figure}

In order to understand how reionization proceeds, it is instructive
to look at the relation between neutral fraction and density. This is 
plotted for the fiducial run in Figure \ref{scatterfid} at 4 different
stages in the evolution of the model. In each panel, we also show a histogram
of particle density values and neutral fractions as well as a scatter plot
of one versus the other. If we go from panel to panel, we can see
that the high density regions are reionized first (the 'inside-out' scenario
also found in Iliev \etal 2006).  For example, when $x_{m}=0.1$, the cloud
of  partially ionized particles to the right of the panel is
centered around $\rho/\left< \rho \right> \sim 10$, whereas at $x_{m}=0.7$
it is centered around $\rho/\left< \rho \right> \sim 3$. Within this
cloud of points, the higher rate of recombinations in high density regions
 means that within the partially ionized volume there 
is a trend for higher neutral fractions
at high densities. This trend is additional to the opposite 
trend for high density regions to be reach this partially ionized state first.
As the universe comes close to being fully
ionized,  $x_{m}=0.97$, particles close to the mean density are the only
ones still fully neutral and the particles with the highest densities
have neutral fractions $< 10^{-6}$. 

\begin{figure}
\centerline{
\psfig{file=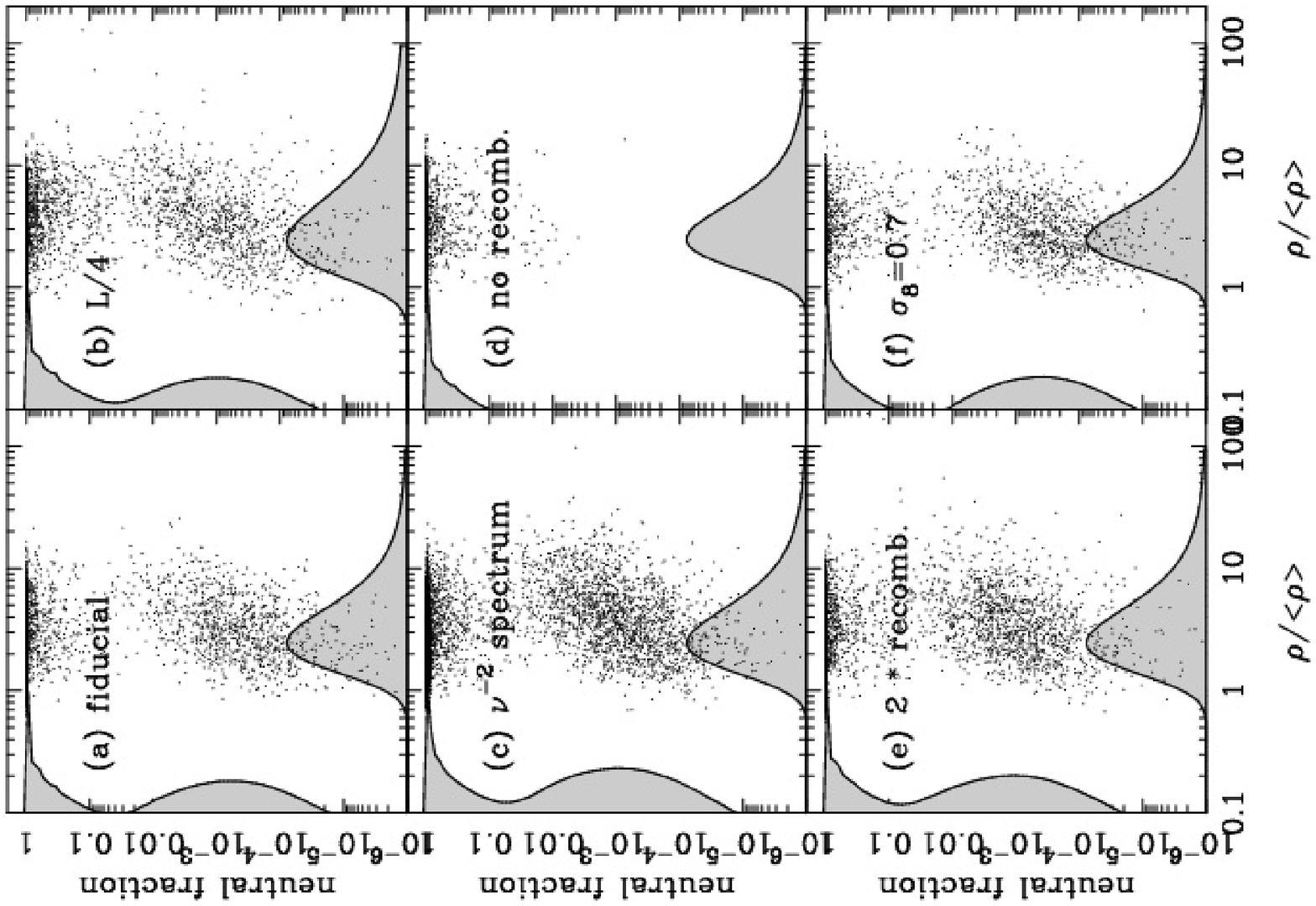,angle=-90.,width=8.0truecm}
}
\caption{
Scatter plot of gas density (in units of the mean) against 
hydrogen neutral
fraction for particles in six different simulation runs (taken from
the 12 in \S 2.3). We show results for all simulations
at the time when the mean mass weighted ionized fraction $x_{m}=0.5$
In each panel, we also show as shaded areas
 histograms of the number of particles
in bins of hydrogen neutral fraction and also histograms binned by
density, $\rho/<\rho>$. The height of each histogram bin is on a log
scale.
\label{scatter_m}
}
\end{figure}

In the next section, we 
examine the morphology of the neutral and ionized regions in order to
investigate this in more detail. For now, we can compare the scatter plots
of neutral fraction and density in different models. We do this
at $x_{m}=0.5$ in Figure \ref{scatter_m}. The absence of a cloud of 
partially ionized points in the ``no recomb'' simulation is expected, as
once particles are ionized, they drop off the bottom of the plot. Also
as expected, the $\nu^{-2}$ spectrum simulation and the ``$2\times$ recomb''
run have a greater density of particles in this region than
the fiducial model.

\section{Morphology}

Just as looking at structure in galaxy redshift surveys 
(e.g., Schectman \etal 1996) revealed
filaments, voids and clusters, the RIS is expected to give rise to 
a complex morphology. In the case of structure in the density field,
reproducing the visual characteristics of observational data was
one of the drivers in searching for the correct theory of structure 
formation, and cosmological N-body simulations are expected
to give rise to a ``cosmic web'' with the same appearance (e.g., Bond
\etal 1996).
Deciding how to view the morphology of RIS is complicated by the 
question of whether to plot the neutral fraction, neutral density,
ionized density or ionized fraction, each of which can lead to a potentially
different impression. Also, unlike pure density fluctuations, which evolve
relatively slowly, the exact time the RIS is plotted as reionization
proceeds can yield very different results. Again, we will choose to 
compare different models at the same value of $x_{m}$.

\begin{figure*}
\centerline{
\psfig{file=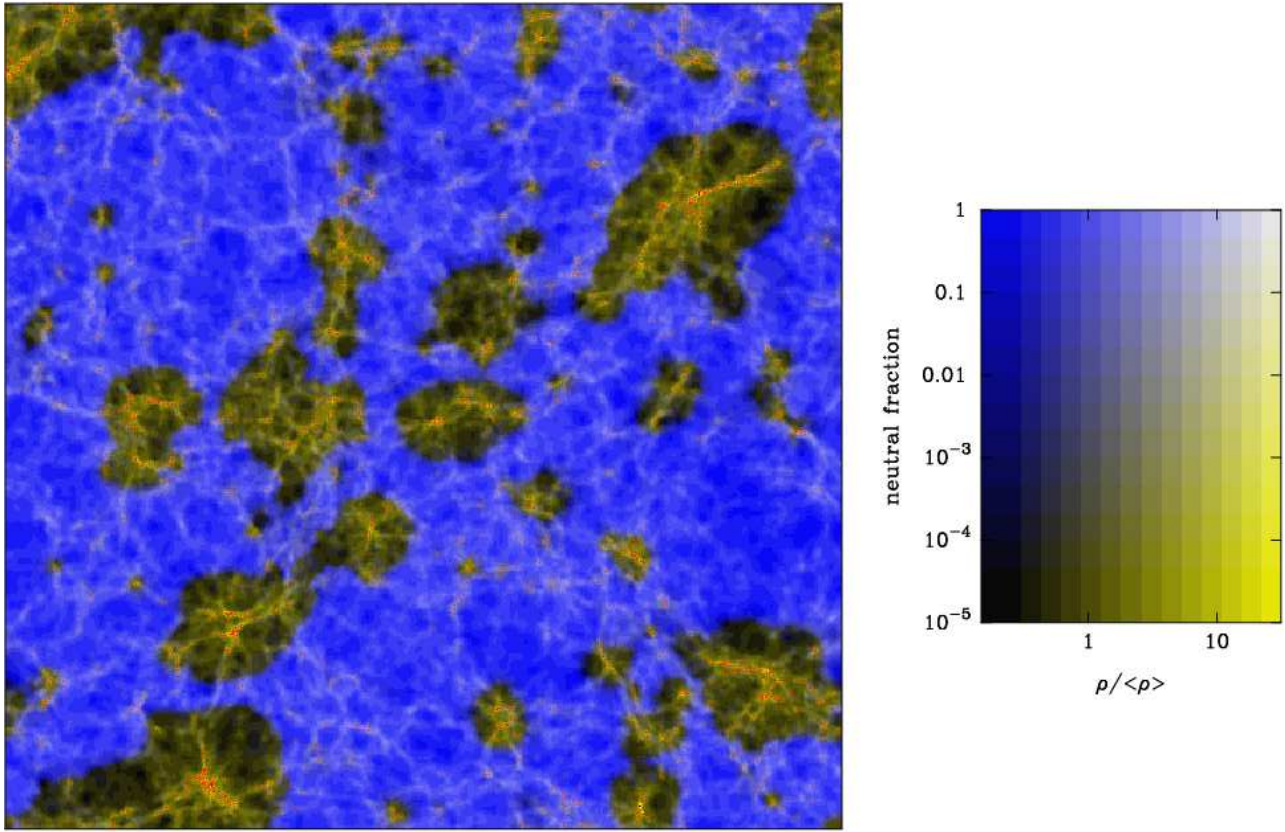,angle=-90.,width=17.0truecm}
}
\caption{A thin (1 $\hmpc$) slice through the fiducial simulation
volume (40 $\hmpc$ wide), at redshift $z=8.5$, when the mean mass weighted
H neutral fraction $x_{m}=0.5$. We show both the density in units of the
mean and the H neutral fraction using a two dimensional color scale. The 
positions of individual sources of ionizing radiation are also shown
as (red) points.
\label{slice_05}
}
\end{figure*}

In Figure \ref{slice_05}, we show a thin (1 $\hmpc$) slice through the 
fiducial simulation volume,
at the time when $x_{m}=0.5$. We use a two dimensional color scale to 
show both the density and the neutral fraction of hydrogen, as well as 
overplotting the positions of the individual sources of radiation,
associated with dark matter halos. It is apparent from the plot that 
low density regions where there are only a few isolated sources have
not yet created noticeable bubbles, but that the highly clustered
regions, associated with filaments and proto clusters have appreciable
Stromgren spheres around them. Iliev \etal (2006)  
 have
measured the size of the bubbles in their simulations by fitting 
spheres into the ionized regions, finding a median
 bubble radius of $\sim 5 \hmpc$ (see their Figure 13) at this late
stage of reionization.
Many other studies, e.g., Shin \etal (2007), Zahn \etal (2007) have
been carried out on the size of bubbles in simulations.
 Visually, our  simulation appears to be broadly consistent with the sizes
found by Iliev \etal,
and we leave to future work detailed statistical characterization of the
size and shape of voids in the neutral hydrogen.  

For now, we will comment on the obvious differences apparent between the
morphology of RIS in the simulation and that which can be seen in the
underlying density field. The RIS has a much higher contrast level,
with the neutral fraction in HII regions being $\sim 10^{5}$
times less than in the rest of the volume.  The edges of HII regions
are consequently much sharper than those of voids in the matter
distribution, lending themselves to easier detection by void-finding 
techniques (e.g., Colberg \etal 2005).

\begin{figure*}
\centerline{
\psfig{file=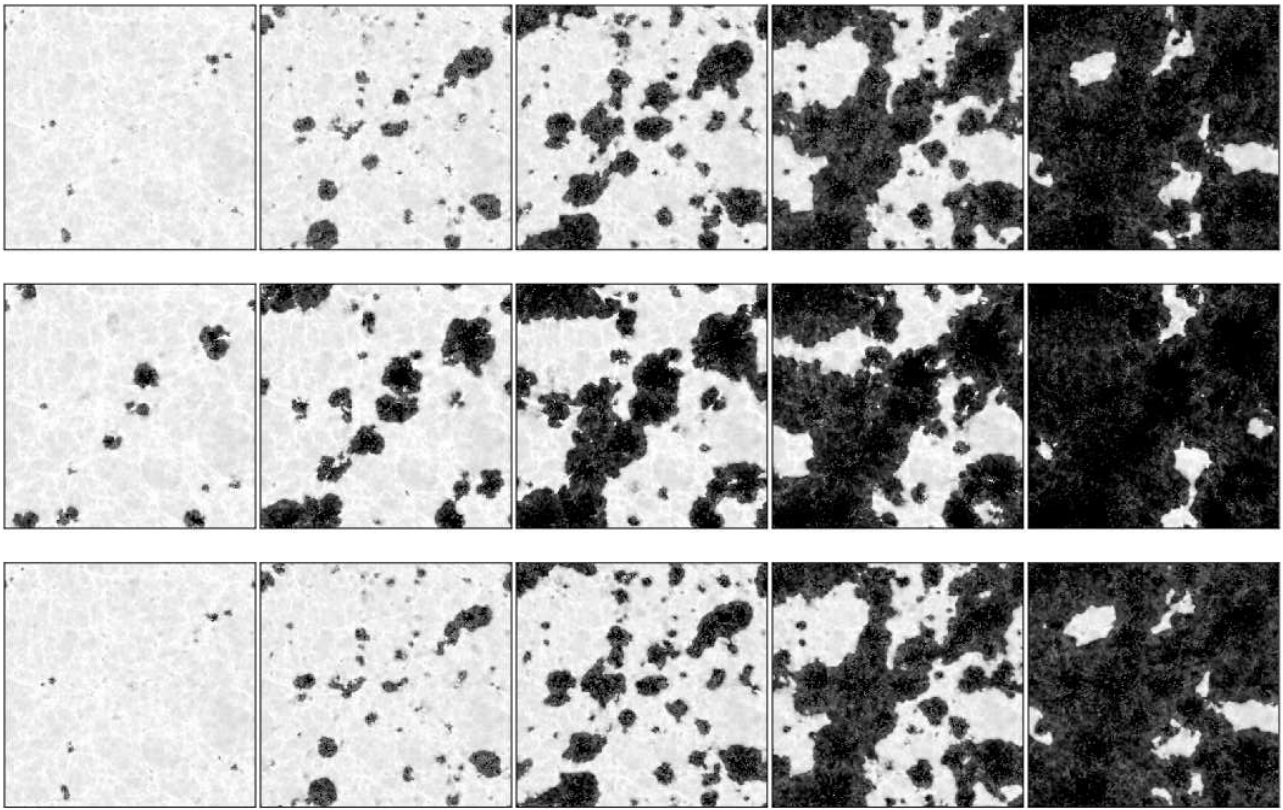,angle=-90.,width=17.0truecm}
}
\caption{
Slices (1 $\hmpc$ thick) 
through the neutral H density field in 3 simulation runs
as a function of mean mass weighted ionized fraction, $x_{m}$.
The top row shows the fiducial run, the middle row
the $M>10^{10}\msun$ run and the bottom row the $M<10^{10}\msun$ run 
(see \S2.3 for full descriptions.) 
In each panel we use a log scale, with light shades
representing high neutral H density. From left to right,
we show results for  $x_{m}=0.1, 0.3, 0.5, 0.7$ and $0.9$.
\label{slice_fbs}
}
\end{figure*}

The matter density field over the redshift range relevant to reioniziation
 evolves little compared to the neutral density field.
 Reionization (say the change from 
 $x_{m}=0.001$ to $x_{m}=0.999$) in these models
takes place over an change in scale factor $a$ of $\sim 2$, and 
as $\Omega_{m}\sim 1$ to good approximation at these redshifts, linear
growth of matter fluctuations occurs by the same factor. Because of this,
most of the change in the morphology and structure of neutral density 
field occurs in the RIS. Plotting a slice through the neutral density
field at different epochs allows us to see this well. In figure
 \ref{slice_fbs}, we show this for 3 different models (Fiducial,
 $M> 10^{10}$, $M< 10^{10}$), at times when $x_{m}$ varied from
0.1 to 0.9 in steps of 0.2. 

McQuinn \etal (2007) have shown
that the large scale morphology of ionized hydrogen bubbles depends
most strongly on $x_{m}$ and the properties of the ionizing sources, 
and is relatively
less affected by the specific subgrid model used to determine small scale 
source suppression and clumping factors.  
In Figure \ref{slice_fbs}, we can see that models 
in the same column (same value of $x_{m}$)
are indeed fairly similar. However, because we use
the mass-weighted ionization fraction $x_{m}$ rather than the volume weighted
fraction $x_{v}$, our conclusions about the similarity of the morphologies
is somewhat different than those of McQuinn \etal (2007).
 For example, the panel
with  $x_{m}=0.5$ (middle panel) for the fiducial model (top row)
 seems to be most similar
to the $x_{m}=0.3$ (2nd panel) for the  $M> 10^{10}$ model (middle row).
This is because in the $M> 10^{10}$ model, the ionized material is 
all concentrated in large bubbles, whereas in the fiducial model there are
many smaller HII regions around the fainter sources which are hard  to see
but which nevertheless account for much of the mass in ionized material.

\begin{figure*}
\centerline{
\psfig{file=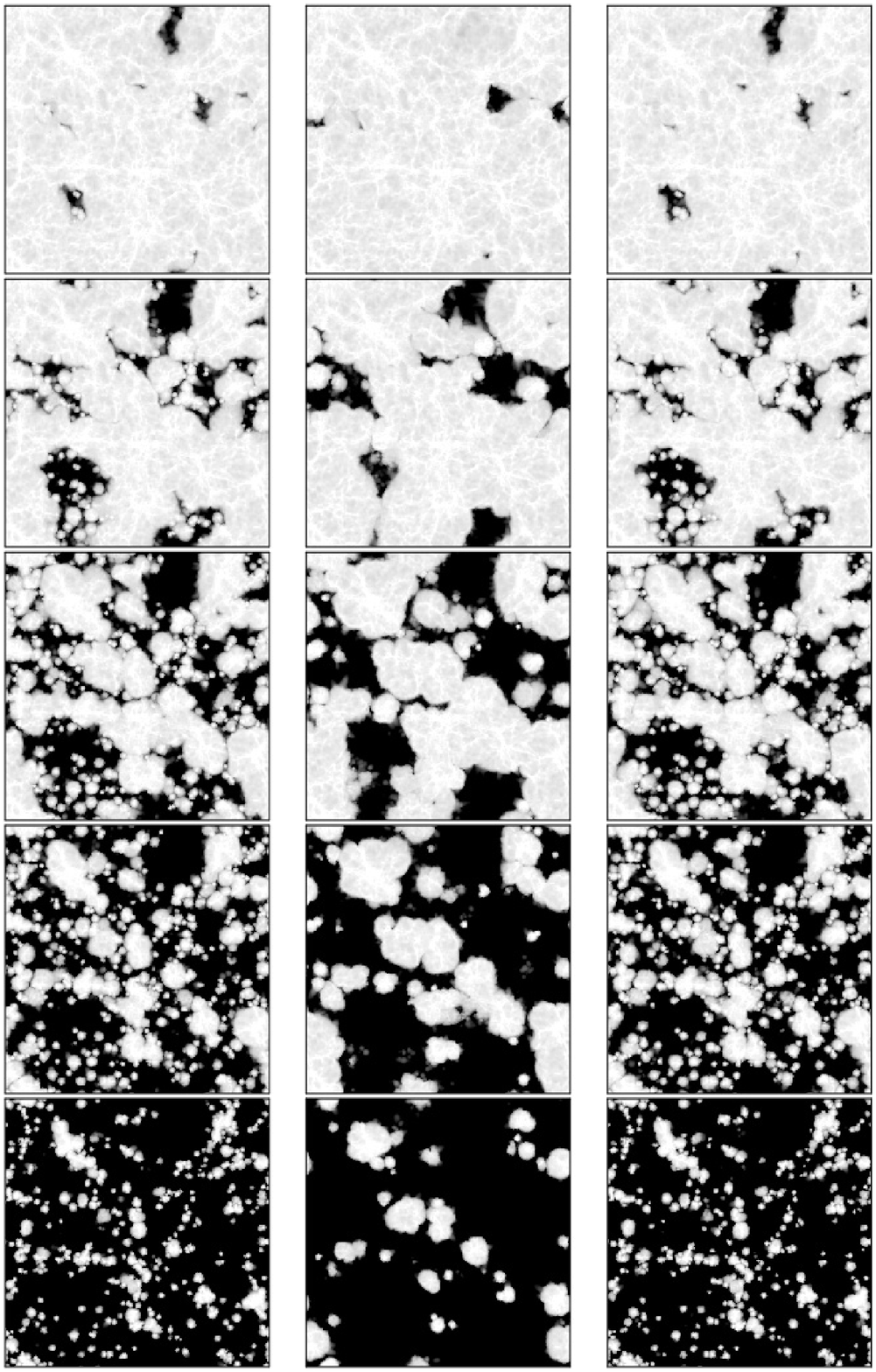,angle=-90.,width=17.0truecm}
}
\caption{
This figure is the same as Fig. \ref{slice_fbs}
except that we show the ionized density: plotted are slices (1 $\hmpc$ thick) 
through the ionized H density field in 3 simulation runs
as a function of mean mass weighted ionized fraction, $x_{m}$.
The top row shows the fiducial run, the middle row
the $M>10^{10}\msun$ run and the bottom row the $M<10^{10}\msun$ run 
(see \S2.3 for full descriptions.) 
In each panel we use a log scale, with light shades
representing high ionized H density. From left to right,
we show results for  $x_{m}=0.1, 0.3, 0.5, 0.7$ and $0.9$.
\label{slice_fbsi}
}
\end{figure*}

A good way to see that this is the case is to refer to Figure
\ref{slice_fbsi},
which shows the same slices through the same models, but this time
plotting the ionized density rather than neutral density. The HII 
regions around the fainter sources
are clearly seen in the top and bottom rows. The early HII regions appear
sharper and perhaps more spherical when seen directly in terms of their ionized
density, rather than as shadows in the neutral hydrogen plot 
(Figure \ref{slice_fbs}). This is understandable because of the fact that
the edges of the bubbles and the totality of bubbles that are
smaller than the slice thickness (1 $\hmpc$) in size
 will be somewhat obscured in  Figure \ref{slice_fbs} by neutral 
hydrogen that lies in front or behind the bubble but is still in the slice. 

In Figures \ref{slice_fbsi} and \ref{slice_fbs} we see little difference
between the $M < 10^{10}\msun$
 model and the fiducial model, indicating that the
absence of the most massive halos does not greatly affect the morphology,
 as long as we compare at the same value of $x_{m}$. If we look at the 
panels at the farthest right of these plots, the end stages of reionization
($x_{m}=0.9$),
we can see that the models with small halos allowed do have HI
remnants with more ragged edges and more small scale structure apparent
in them than the $M > 10^{10} \msun$
 model. In future work, it would be interesting to
investigate the mass function of the disconnected HI remnants present 
at these times as they may have constraining power (as well as
being likely detectable in 21cm emission. The $x_{m}$ value at which 
the ionized regions percolate seems likely to depend on the source model
also, as for example the $x_{m}=0.5$ panel for the 
$M > 10^{10} \msun$ model consists
less of disconnected HII regions than the other two panels. This makes sense,
as the ionized material is more clustered, being closer to 
the more massive haloes.

\begin{figure*}
\centerline{
\psfig{file=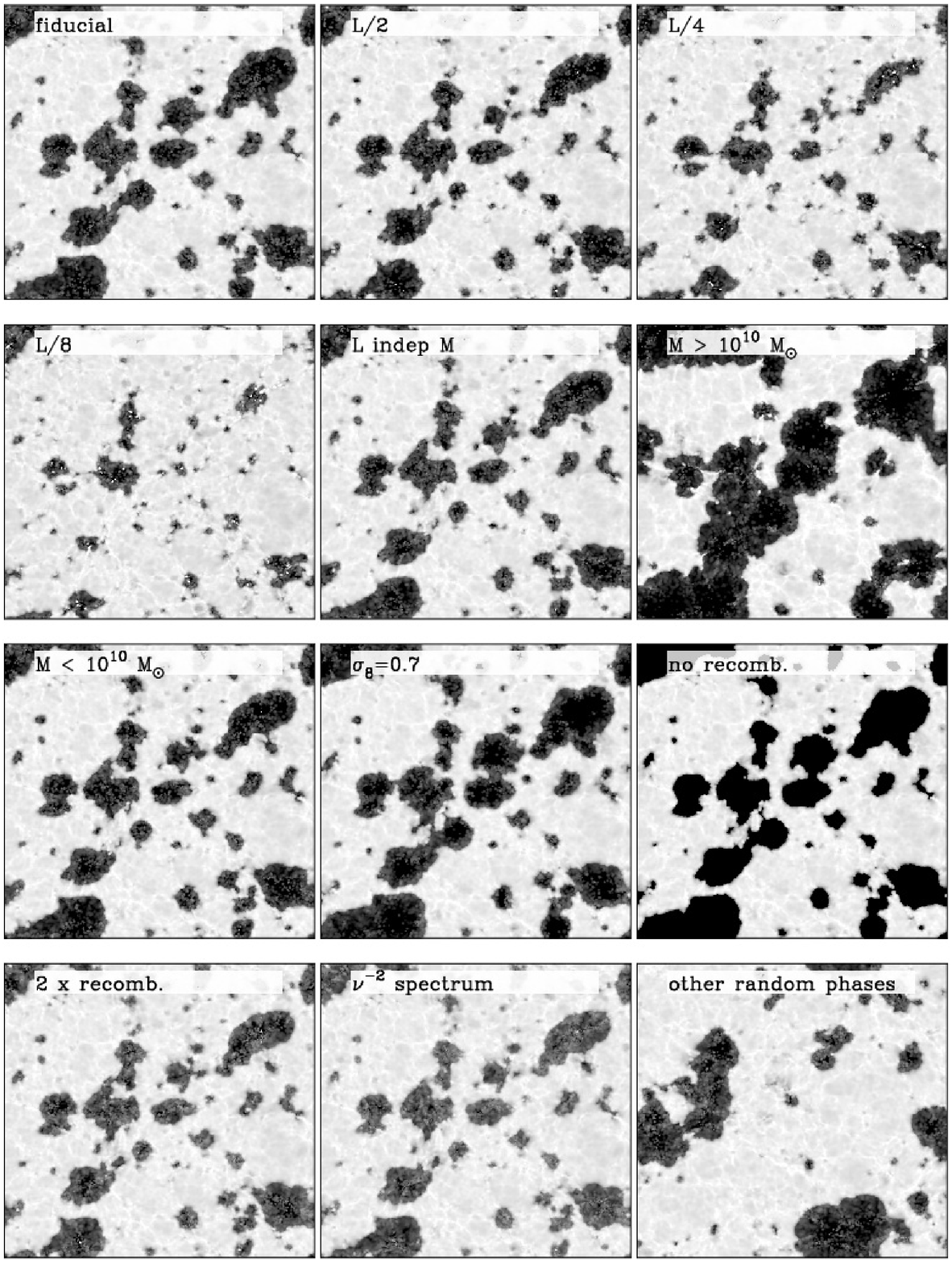,angle=0.,width=14.0truecm}
}
\caption{
Slices (1 $\hmpc$ thick) 
through the neutral H density field in the 12 simulation runs
described in \S 2.3. We show all results at output times that 
correspond to a mean mass weighted ionized fraction, $x_{m}=0.5$.
The density is shown on a log scale (light colours for higher neutral
density). Regions which are completely black generally have a neutral
fraction $< 10^{-6}$.
\label{slice_m}
}
\end{figure*}

This $x_{m}=0.5$ epoch is therefore a good one at which 
to compare the morphology of the other models also. In Figure \ref{slice_m},
 we show the same thin slice through the neutral density field for the 12
different models. The panel most different from the others is not 
suprisingly that for the ``other random phases'' model, for which the slice
intersects a fairly spherical bubble (at the bottom of the plot) and 
a large region which has little sign of reionization (in the middle). 
Comparing this panel to the fiducial model, it might seem as though the
latter is closer to the percolation stage, although they both have 
the same  $x_{m}=0.5$. In the later stages of reionization, when
the ionized structures are a substantial fraction of the box size, we 
must naturally be careful in our interpretation of the morphology,
due to the effects of cosmic variance and finite volume. As
stressed by Iliev \etal (2006), and others,
large boxes are necessary to capture these processes,
particularly at the later epochs.

Looking systematically at the different panels of Figure \ref{slice_m},
 we can see that in the fiducial, L/2, L/4 and L/8 models as the luminosity
per halo atom is reduced, the visual extent of the structures
becomes smaller and smaller. From Figure \ref{recpath} we can see that
in these models, the number of  recombinations becomes progressively
higher as we reduce the luminosity, and although one would expect that this
would be compensated for in the morphology by the fact that we
compare all models at the same $x_{m}$, there is in fact still a large 
difference. We can therefore state that changing the luminosity of 
sources at fixed  $x_{m}$ does affect the visual morphology quite strongly.
For the L/8 model, much of the ionized hydrogen is again hidden from 
view in the  strongly neutral regions, as happened in the comparison
of the fiducial and $M > 10^{10}\msun$
 models. From a rough visual comparison of 
the different columns of Figure \ref{slice_fbs} and models L/2-L/8 in Figure 
\ref{slice_m} it seems as though scaling the luminosity of sources
by a factor of $\sim 4$ changes the visual morphology of the HI density
field in a similar fashion to changing $x_{m}$ by $\sim0.2$. Of course 
this will not hold for the morphology of the ionized regions, and we
will investigate this statistically when we look at the correlation function
in the next section.

\section{Autocorrelation function}

Quantitative comparisons of the structure in the various models can
be carried out by looking at the autocorrelation function, $\xi_{r}$. 
We will focus on $\xi(r)$ measured for the neutral density distribution,
although we will briefly examine  $\xi(r)$  for the ionized density field.
We note that the Fourier transform of  $\xi(r)$, the power spectrum 
$P(k)$ of fluctuations has been studied in reionization models by 
many authors. We compute $\xi(r)$ for the gas density directly 
from the particle positions in the simulation, and for 
 $\xi_{HI}(r)$ we weight each particle by its neutral fraction: 
\begin{equation}
\xi_{HI}(r)= \frac{\sum_{N_{p}} (x_{HI})_{j} (x_{HI})_{k}}
{N_{p,e}\left<x_{HI}\right>^{2}}-1,
\end{equation}
where $N_{p}$ is the number of pairs of particles in a bin 
centered on $r$, $N_{p,e}$ is the expected number for a random
distribution, $(x_{HI})_{j}$ and $(x_{HI})_{k}$ are
the neutral fractions of particles in a pair, and $\left<x_{HI}\right>$
is the mean neutral fraction by mass.
 As with our examination of
morphology of the neutral density, we will compute  $\xi(r)$ for different
simulation outputs chosen by their mass weighted ionized fraction, $x_{m}$.

\subsection{Resolution and boxsize tests}

\begin{figure}
\centerline{
\psfig{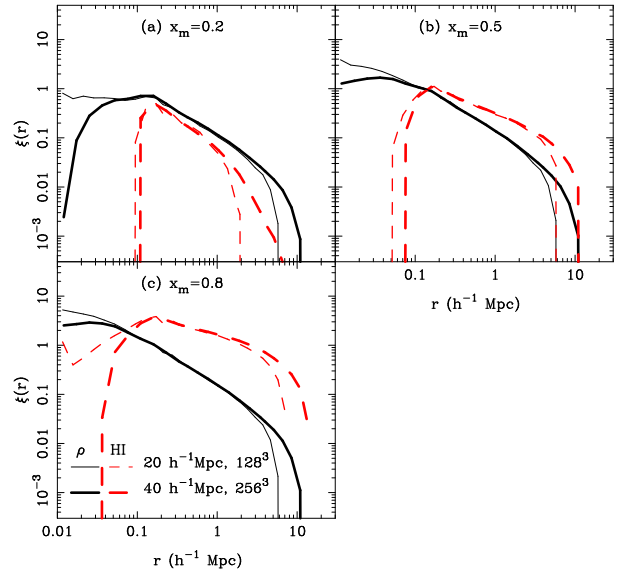}
}
\caption{
The effect of boxsize on $\xi$: We show results
for two simulations, the fiducial run (which has
a $40 \hmpc$ cubical simulation volume, bold lines) and another with identical
mass resolution and other parameters but in a $20 \hmpc$ box (thin lines.)
Panels (a)-(c) show results for different outputs, labelled
by the value of the 
mean mass weighted ionized fraction, $x_{m}$. As dashed and solid lines we
show the autocorrelation function of the HI density and the gas density
respectively.
\label{xi_boxsize}
}
\end{figure}

In order to test the range of validity of our $\xi(r)$ results, we carry out
several resolution and boxsize tests.  Because the bubble like structures
which overlap during the end stages of reionization occupy a large volume,
 one would expect that the simulation box size may have a strong effect on
our results. We therefore compute 
 $\xi(r)$ for the neutral gas and the total gas density for two simulations
with different box sizes (we vary the box side length by a factor of 2)
 but the same mass and spatial resolution (this is kept the same
as our fidcucial model). The 
results are show in Figure \ref{xi_boxsize}, where the 3 panels  show
$\xi(r)_{\rho}$ and
$\xi(r)_{HI}$ 
 for at different stages of reionization (parametrized by $x_{m}$)
in the fiducial model.

We can see that the $\xi(r)_{\rho}$ curves  are extremely
similar over the range $0.05 \hmpc < r < 4 \hmpc$ for the 3 values of
$x_{m}$. In particular, a power law fit to the curves over this range
 gives virtually identical parameters. This is a very good thing, because
it shows that no non-linear gravitational
 mode coupling has taken place with large
scale density modes of order the box size. This is one advantage of working
at these high redshifts ($z \sim 8$ for $x_{m}=0.8$ in this case),
where we can see that even a $20 \hmpc$ box is large enough to study
gravitational clustering over this range of length scales.
The correlation function of the neutral gas density has a 
broadly similar behaviour. The position of the
 break in the powerlaw form of $\xi(r)$
is due to the finite size of the simulation volume and  so we will 
restrict our analysis of  $\xi(r)$ to smaller scales.

Interestingly, the panel with results closest to the end
of reionization ($x_{m}$) does not show any greater disagreement 
for large $r$ for  $\xi(r)_{HI}$  than for $\xi(r)_{\rho}$. This means that
at least over this limited range of scales, our  $\xi_{r}$ measurements
will also be reliable for the neutral density. On the smaller scales 
($r< 0.1 \hmpc$, there appears to be a rapid drop off in  $\xi_{r}$
for the neutral density for both boxsizes. We shall see below when we
consider mass and spatial resolution that our results will not be
useful below these scales in any case.

\begin{figure}
\centerline{
\psfig{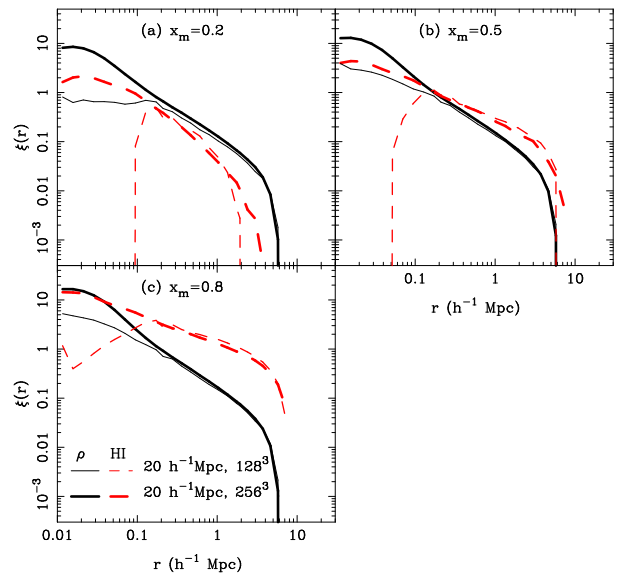}
}
\caption{
The effect of mass and spatial resolution on $\xi$: We show results
for two simulations, both in a
$20 \hmpc$ cubical volume.
One has the same mass and spatial resolution as the fiducial run 
(meaning $128^{3}$ gas particles in this volume, thin lines) and the
other has 8 times better mass resolution and two
times better spatial resolution (meaning $256^{3}$ gas
 particles in this volume, thick lines) 
Panels (a)-(c) show results for different outputs, labelled
by the value of the 
mean mass weighted ionized fraction, $x_{m}$. As dashed and solid lines we
show the autocorrelation function of the HI density and the gas density
respectively.
\label{xi_res}
}
\end{figure}

For our next test, we keep the box size fixed at $20 \hmpc$ but vary the 
particle number (and hence mass resolution) by a factor of 8.
We also vary the spatial (force) resolution by a factor of 2. The coarser
mass/spatial resolution is the one we use in our fiducial case (only with a 
$40 \hmpc$ box). The dark matter halos which we use to place our sources
of radiation have the same lower mass cutoff 
$1.6\times10^{9}$ $\msun$ in both runs,
which corresponds to 8 times fewer particles in the low resolution case.
The results for  $\xi(r)_{\rho}$
and  $\xi(r)_{HI}$ are shown in Figure \ref{xi_res}, again for different
values of $x_{m}$. The  $\xi(r)_{\rho}$ curves show good agreement on 
large scales (the two simulations were run with the same initial 
phases for the density field). On scales  $r< 0.15 \hmpc$, the two curves
diverge, indicating the effects of mass and spatial resolution on
the gravitational evolution of the gas density field.  We note that on
scales comparable to this the clustering in the gas density will be
influenced by cooling and star formation, which we do not include
here in any case. The $\xi(r)_{HI}$ correlation functions also agree well
down to  $r< 0.15 \hmpc$, when they diverge even more sharply. We notice
that as will the boxsize test, the worst disagreement on large scales
occurs rather unexpectedly for the early stages of reionization
($x_{m}=0.2$).

\begin{figure}
\centerline{
\psfig{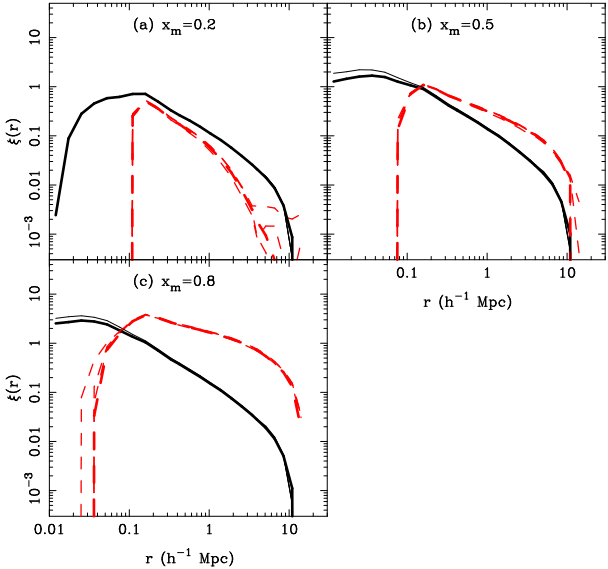}
}
\caption{
The effect of the photon packet size parameter, $c_{f}$ (see \S2.2)
on $\xi$. We show results
for 4 simulations, the fiducial run (which has
$c_{f}=0.3$) and 4 other runs with
$c_{f}=3,1 $ and $0.1$.
The fiducial run is shown as the thickest line and the others as thin lines.
Smaller values of $c_{f}$ imply a larger total number of photon packets.
Panels (a)-(c) show results for different outputs, labelled
by the value of the 
mean mass weighted ionized fraction, $x_{m}$. As dashed and solid lines we
show the autocorrelation function of the HI density and the gas density
respectively.
\label{xi_cf}
}
\end{figure}

The final simulation parameter which we vary is $c_{f}$, the 
maximum number of simulation particles which can be ionized by a single
photon packet. This parameter (see Section 2.2 for more details)
is inversely proportional to the total number of photon packets used
to carry out the Monte Carlo RT.  For larger values of $c_{f}$, the
radiation field will not be as smooth, and there will be more
shot noise in the neutral density field.
Our fiducial value of $c_{f}=0.3$, and in Figure \ref{xi_cf} we show
what happens when this is varied from $c_{f}=3$ to $c_{f}=0.1$,
with all other simulation parameters the same as our fiducial model.
The   $\xi(r)_{\rho}$ curves are almost identical, as might be expected,
with the small differences due to the fact that reionization occurs
at slightly different times in the different $c_{f}$ runs. The curves
for $\xi(r)_{HI}$ are also very similar, and there is
no apparent systematic effect even when $c_{f}$ is made 10 times
larger than our fiducial value. At least as far as the correlation
function is concerned, we have therefore converged to
stable results with our fiducial number of photon packets.

The tests in this section have therefore revealed that our results
for $\xi_{\rho}(r)$ should be reliable over at least the scales 
$0.15 \hmpc < r < 4 \hmpc$. We will concentrate on this range
in our analysis, for example
looking at the power law nature of  $\xi_{\rho}(r)$. For
looking at larger scales,
approaching the scale of bubbles at the time of percolation, larger 
simulation volumes should be run in the future. We note that because
$\xi_{\rho}(r)$ for a $20 \hmpc$ volume converged with a larger box on 
scales below  $r < 4 \hmpc$  it is probably safe to assume that we can draw
information from our fiducial volume  ($40 \hmpc$ box) on scales
up to $r\sim 8\hmpc$.

\subsection{The evolution of $\xi(r)$}

\begin{figure}
\centerline{
\psfig{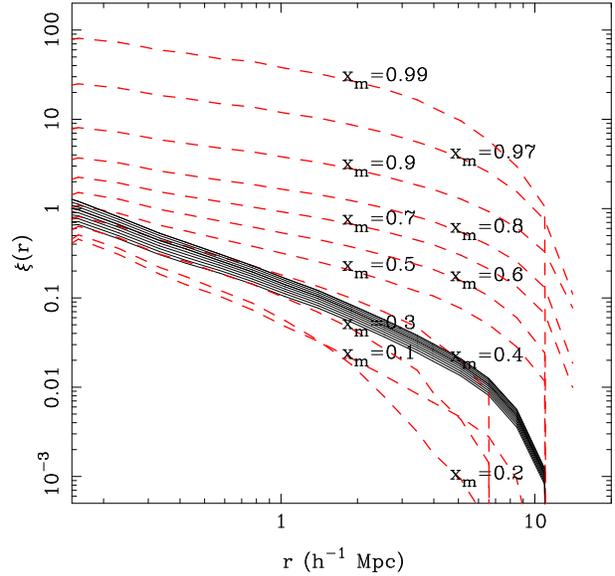}
}
\caption{
The autocorrelation function, $\xi(r)$ for the gas density field
(solid lines) and the HI density field (dashed lines) in the
fiducial simulation (see \S 2.3.) We show results for 11 different output
times, for which we have labelled the HI curves with the mean mass
weighted ionized fraction at that time, $x_{m}$.
\label{xifid}
}
\end{figure}

We show $\xi(r)$ for the fiducial model in figure \ref{xifid},
for values of the ionized fractions $x_{m}$ ranging from 0.1 to 0.99,
 which corresponds to a redshift range of $z=10.2$ to $z=7.6$.
 Before reionization
starts, when $x_{m}=0$, $\xi_{\rho}(r)$ and $\xi_{HI}(r)$ are identical,
by definition. However, once $x_{m}$ has reached 0.1, one can see that 
$\xi_{HI}(r)$ is somewhat lower than  $\xi_{\rho}(r)$, by a factor of $\sim
0.7$, but has the same shape, a power law:
\begin{equation}
\xi(r)=(r/r_{0})^{-\gamma},
\end{equation} 
with slope $\gamma \sim 1.5$ on scales $r \simlt 7 \hmpc$ 
and a gentle break above it (at a scale dictated
by the finite box volume). We will explore power law fits to $\xi(r)$ in 
section 5.4 below. For now, we explore qualitatively the behaviour of 
$\xi_{\rho}(r)$ as reionization proceeds.
We note that this behaviour (a decrease and then an increase
in the amplitude of clustering) has been seen in analytic
calculations of the 21 cm brightness correlations 
(e.g., Wyithe \& Morales, 2007)

As $x_{m}$ increases, $\xi_{\rho}(r)$ exhibits the usual linear growth,
with the amplitude increasing as expected under gravitational instability,
and the shape remaning constant. As can be seen from the solid lines in 
Figure \ref{xifid}, this gravitational amplification of structure is
very small (a factor $\sim 1.5^{2}$) over the time of reionization. 
$\xi_{HI}(r)$ on the other hand shows dramatic growth, by a factor 
100 or more over the same interval, as we would predict for example
by examining the morphology of structure in the HI density field in 
Figure \ref{slice_fbs}. The initial stage of reionization affects primarily
the high density regions around sources of radiation. Removing their
contribution from the clustering of HI leads to an antibias on 
all scales in $\xi_{HI}(r)$ with respect to  $\xi_{\rho}(r)$. After this,
 the effect of removing highly clustered neutral regions competes with
the amplifying effect of RIS, with the latter winning 
after $x_{m}\sim 0.35$, raising $\xi_{HI}(r)$ above  $\xi_{\rho}(r)$
at this point. As the RIS grows in scale, the shape of $\xi_{HI}(r)$
begins to change, with the slope of the powerlaw region becoming flatter,
 reaching $\gamma \sim 0.5$ when $x_{m}=0.99$.

\begin{figure}
\centerline{
\psfig{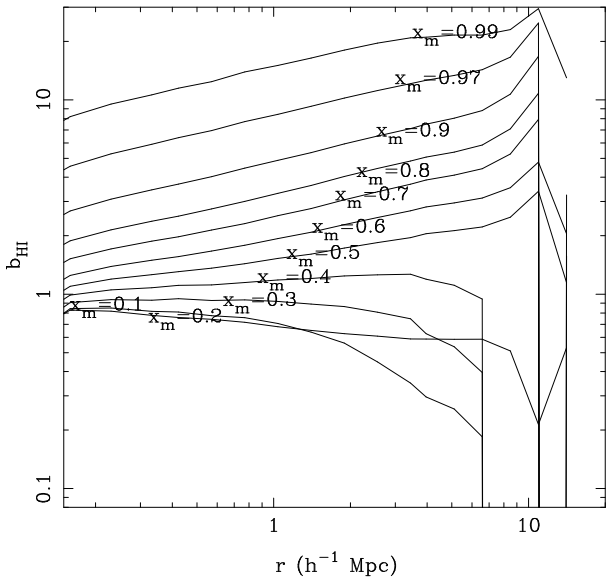}
}
\caption{The HI bias factor, b$_{HI}$ as a function of 
scale, $r$ for the fiducial simulation. The factor b$_{HI}(r)$
is given by Equation \ref{beq}
\label{bfid}
}
\end{figure}

The relationship between $\xi_{HI}(r)$ and  $\xi_{\rho}(r)$ can 
be examined by plotting the bias as a function of scale,
defined by 
\begin{equation}
b_{HI}(r)= \left[\frac{\xi_{HI}(r)}{\xi_{\rho}(r)}\right]^{1/2}
\label{beq}
\end{equation}.
This quantity is plotted in Figure \ref{bfid} for the fiducial model,
 where it can be seen that $b_{HI}(r)$ is approximately constant
with scale for $r\simlt 5 \hmpc$ for $x_{m} < 0.35$, with a value less than 1
for low values of $x_{m}$. After this,  $b_{HI}(r)$ takes on a positive
power law slope. If  $\xi_{HI}(r)$ and  $\xi_{\rho}(r)$
are fit by powerlaws, with slopes $\gamma_{HI}$ and $\gamma_{\rho}$,
then we expect the slope of a power law fit to 
$b_{HI}$,
\begin{equation}
b_{HI}(r)=(r/r_{0})^{-\gamma_{b}},
\end{equation}
to  give $\gamma_{b}=\frac{1}{2}(\gamma_{HI}-\gamma_{\rho})$. If we examine
$b_{HI}$ for $x_{m}=0.99$, we find a slope 
$\gamma_{b}=\frac{1}{2}(0.5-1.5)=-0.5$, as expected. We can see that this
slope is reached quite rapidly after  $x_{m}$ reaches $\sim0.5$ and does not 
appear to evolve much after this. We shall explore this further below.
The behaviour of $b_{HI}$ in this strongly ionized regime, is not constant
with scale, indicating that during the late stages of reionization, 21cm
tomography measurements (if they were possible on these small scales,
which is observationally difficult) would not result in a straightforward 
measurement of the underlying gas or dark matter clustering. During
the earlier stages, however, the (anti)bias is linear, although its amplitude
may be related to reionization process in a complex way.

\begin{figure}
\centerline{
\psfig{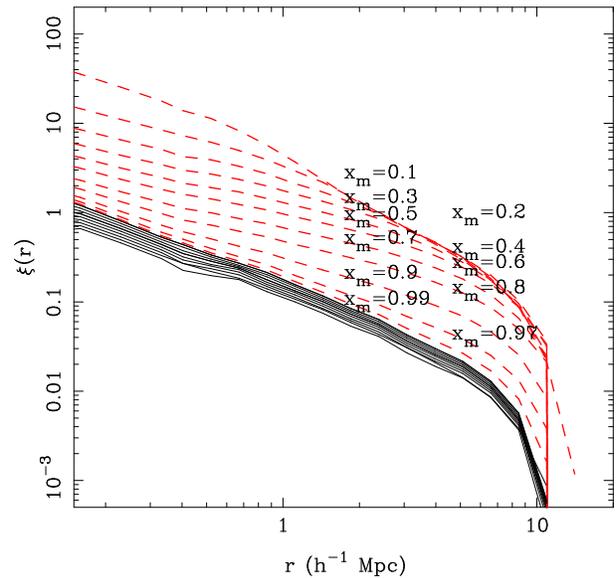}
}
\caption{Clustering of ionized hydrogen:
we show the autocorrelation function, $\xi(r)$ for 
ionized H density field (dashed lines) and
the gas density field
(solid lines) in the
fiducial simulation (see \S 2.3.) We show results for 11 different output
times, for which we have labelled the ionized H curves with the mean mass
weighted ionized fraction at that time, $x_{m}$.
\label{ionxifid}
}
\end{figure}

If instead we compute $\xi_{H^{+}}(r)$, the correlation function of
the ionized hydrogren (e.g., as plotted in Figure \ref{ionxifid}),
 we find that the situation is simpler. This is
shown for the fiducial model for different $x_{m}$ values in 
Figure \ref{ionxifid}. At the highest values of $x_{m}$, we obviously
expect $\xi_{H^{+}}(r) \sim \xi_{\rho}(r) $, which has the characteristic
power law behaviour for $r \simlt 8 \hmpc$. At earlier times, $\xi_{H^{+}}(r)
$ has the same slope, but a larger amplitude, indictating that the
ionized regions trace a biased subset of the density distribution, around
peaks in the density field (see e.g., Kaiser, 1984.)
When $x_{m}=0.1$ (at $z=10.2$), 
the value of $r_{0}$ is $\sim 3 \hmpc$, (compare
to $5 \hmpc$ for $L_{*}$ galaxies at the present
day, Zehavi \etal 2005).
We can see from Figure \ref{ionxifid} that  $\xi_{H^{+}}(r) $ breaks
from a powerlaw at around $r\sim 1\hmpc$, which, from looking at 
Figure \ref{slicefbs} corresponds approximately 
to the size of ionized
bubbles at this early epoch. 

\subsection{$\xi(r)$ for the different models of reionization}

We have seen that $\xi_{HI}(r)$ in the fiducial model has a similar shape
to that which arises under gravitational instability. We now investigate
 $\xi_{HI}(r)$ in the other reionization models in order to see if there
is a dependence on the properties of the sources of radiation or
the physics of reionization. We again compute  $\xi_{HI}(r)$ at 
output times corresponding to different specific values of $x_{m}$.

\begin{figure*}
\centerline{
\psfig{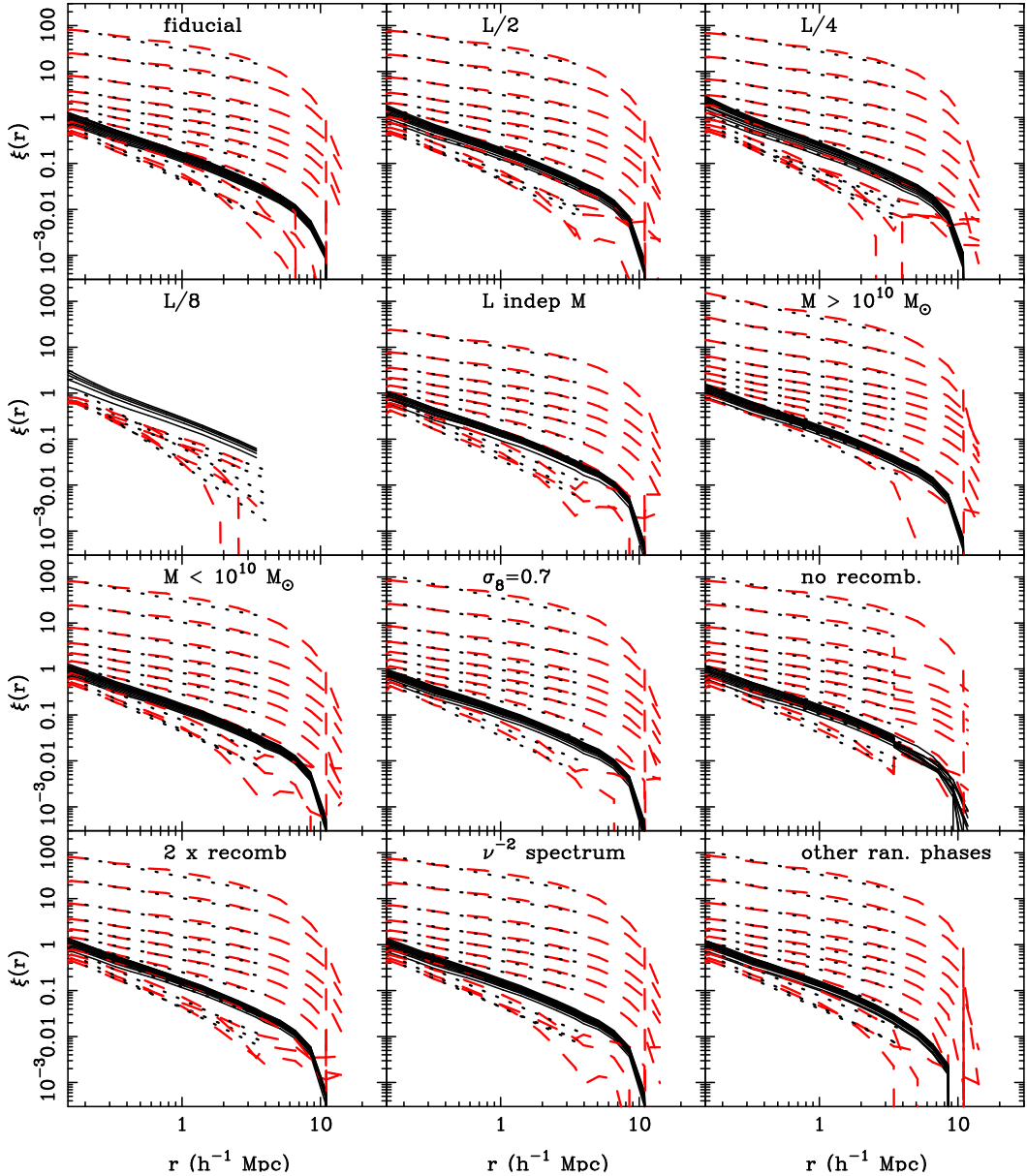}
}
\caption{
The autocorrelation function $\xi$ for all 12 models
described in \S2.3. The dashed lines are $\xi(r)$ for the
neutral HI and solid lines for the gas density.
 The dotted lines show power law fits to the
HI $\xi(r)$ for the region $4.0 \hmpc > r > 0.2 \hmpc$. For all panels
except those for
the ``L/8 ``and ``L indep M'' models, we show results for the same values 
of the
mass weighted mean ionized fraction as in Figure \ref{xifid}, i.e.
$x_{m}=0.1-0.9$ in steps of 0.1 and then  $x_{m}=0.97$ and $x_{m}=0.99$.
 For the ``L indep M'' models, the $x_{m}=0.99$ lines are not shown.
\label{xi_m}
}
\end{figure*}

 The results for our 12 models are shown in Figure \ref{xi_m}, along with 
power-law fits (described more fully in Section 5.4 below) 
to the region $4.0 \hmpc > r > 0.2 \hmpc$.
 It is apparent
that all models display behaviour broadly similar to the fiducial case,
with a form reasonably approximate to a power law over $\sim 1.5$ decades
in scale. As indicated by our tests above, the break on scales $r \sim
7 \hmpc$ is largely caused by the finite box volume and is expected to be
similar in all cases.  All models have a certain level of antibias
between  $\xi_{HI}(r)$  and  $\xi_{\rho}(r)$ at first, with the
most extreme antibias being reached in the ``L/8'' model. The ``$M>10^{10}
\msun$'' model has only a small level of antibias, indicating that the
large bubbles formed rapidly by the very luminous sources quickly modulate
the HI field.

The correlation function becomes shallower
and its  amplitude increases
dramatically for all models towards the end of reioniziation. 
When $x_{m}=0.99$, the models all have a very similar amplitude and a
shallow power law slope $\gamma \sim 0.5$. It should be noted that model 
``L/8'' and ``L indep M'' do not have outputs for the very end stages of 
reionization. The latter has very similar behaviour to the fiducial case,
and the former is rather different, as we shall see when we consider
the evolution of the power fit parameters.

Overall one can see from the panels in  Figure \ref{xi_m} that the RIS leads
in all models to a strong increase in  $\xi_{HI}(r)$ over the course of
reionization. Unlike linear growth of fluctuations under GI, the
slope the correlations changes, reaching a similar shallow value in 
all models. No obvious features with any
particular physical are created in  $\xi_{HI}(r)$ in any of
our variations of the reionization scenario. Despite the 
relatively wide
variety of models employed, the differences in $\xi_{HI}(r)$
are subtle, and will be explored in $r_{0}-\gamma$ space below.

\subsection{Power law fits}

The galaxy-galaxy correlation function has only small deviations
from power law behaviour over $\sim 3$ decades in length scale 
(e.g., Peebles, 1980, and Zehavi \etal 2002),
 despite the undoubted 
complexities of galaxy formation. The dark matter correlation
function in simulations also exhibits this behaviour 
(e.g., Jenkins \etal 1998, Kravtsov \etal 2004.)
It is therefore not unreasonable to 
expect that this type of scale invariance might also be created by 
RIS. We have seen from Figure \ref{xi_m}  that this is indeed the case.
In this subsection we examine the power law behaviour of $\xi_{HI}(r)$
more quantitatively. 

\begin{figure}
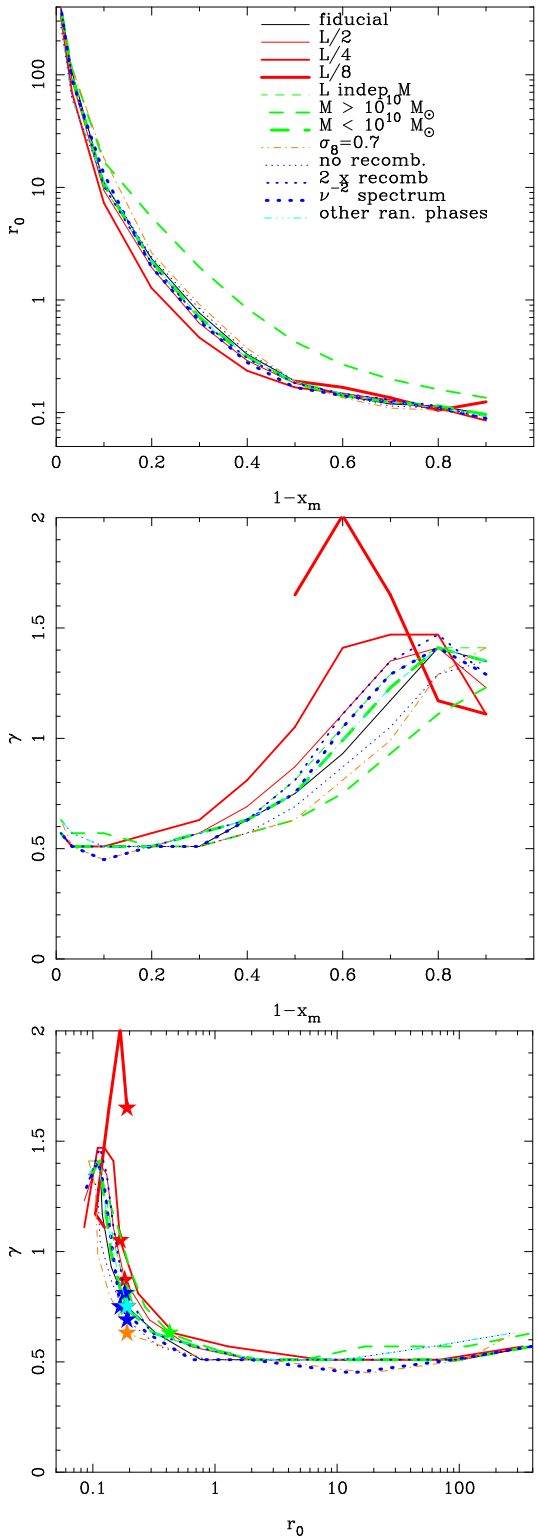

\begin{center}
\begin{minipage}[t]{2.3in}
\psfig{file=fitxi1.ps,angle=-90.,width=7.0truecm}
\psfig{file=fitxi2.ps,angle=-90.,width=7.0truecm}
\psfig{file=fitxi3.ps,angle=-90.,width=7.0truecm}
\end{minipage}
\caption{Parameters for power law fits to the autocorrelation
function of the HI density in all 12 models described in \S2.3.
Top panel: correlation length $r_{0}$ as a function of mass weighted
ionized H density $x_{m}$. Middle panel: slope $\gamma$ as a function
of $x_{m}$. Bottom panel:  $r_{0}$ vs. $\gamma$.
\label{fitxi}
}
\end{center}
\end{figure}

We fit power laws to the region $4.0 \hmpc > r > 0.2 \hmpc$,
 based on the resolution and boxsize tests of Section 5.1.
The power-law fits were carried out assuming Poisson errors on
the $\xi_{HI}(r)$ points, so 
that the error is $\sigma= (1+\xi_{HI}(r))/\sqrt(N_{p})$, where $N_{p}$ is the 
number of  pairs of particles in a bin. We have tried changing $N_{p}$ to 
include only pairs of particles above a threshold neutral fraction
(e.g., $x_{HI}=0.5$), but this has a negible effect on the fit parameters.
Also, changing the fitting region to $1.0 \hmpc > r > 0.2 \hmpc$
does not change any of our conclusions below.

In the top 2 panels of Figure \ref{fitxi}, we show the dependence
of the power law fit parameters $r_{0}$ and $\gamma$ on $x_{m}$
for our different models. Looking at $r_{0}$ first, we can see that
there is a strong dependence  on $x_{m}$.
The models all follow a trend roughly equivalent (within a factor of
2 in $r_{0}$ for all but one model)
 to $\log_{10} r_{0} \sim  0.06/(1.-x_{m})^{2}$, 
with the $M> 10^{10}\msun$ model being the most extreme outlier.
 The 
fact that the lines for nearly all the different models are clustered 
together so
 tightly is rather surprising, and appears to be an
indication that $x_{m}$
plays a dominant role setting $r_{0}$. At least in this fashion,
the neutral fraction governs the structure in the neutral hydrgogen
density field. The visual morphology of different models with
 the same $x_{m}$ values (e.g., Figure \ref{slice_m})
 did appear to be noticeably
different, rather more than one would expect give the tight locus
of $r_{0}-x_{m}$ curves. Given the visual impression of the HI slices,
however, it is understandable that the model with perhaps the greatest
difference to the others ( the  $M> 10^{10} \msun$ model) is also discrepant
in terms of $r_{0}$. For any value of $x_{m}$, this model has a larger
$r_{0}$ than the others.

The $r_{0}$ differences between models are 
 small, but measurable, as are
the purely gravitational instability based differences (note that 
the models reach different values of $x_{m}$ at different redshifts,
so that the underlying $\xi_{\rho}$ curves will be different.)

We do find a wider variation in the values of the slope of $\xi_{HI}$,
for fixed values of $x_{m}$. This can be seen from the middle panel 
of Figure \ref{fitxi}, where for example $\gamma$ varies from 
1.6 to 0.6 when $x_{m}=0.5$. The fiducial model has $\gamma=0.72$ when
$x_{m}=0.5$. If we look horizontally to find out at what  $x_{m}$ value
the other models have the same slope, we find a range from $1-x_{m}=0.34$
(for the L/4 model) to $1-x_{m}=0.60$ (for the $M> 10^{10}\msun$ model).
All curves (except for the L/8 model which does not get more
than half ionized) do follow the same pattern, with the slope getting
asymptotically flatter as reionization proceeds, all ending up 
with $\gamma \sim 0.5$. The rapidity with which this asymptotic
value is 
reached does vary with the different models, with the $M> 10^{10}\msun$ model
doing this most quickly, as the large-scale modulation
effects of the bright sources tilt the correlation function upwards on
large scales.

The slope of  $\xi_{HI}$ stays approximately constant as $r_{0}$ increases
rapidly during the end stages of reionization, as can be seen clearly from
the bottom panel of Figure \ref{fitxi}, where $r_{0}-\gamma$ curves are
plotted. In the region to the left of the plot, clustering of
HI in all models is dominated by the clustering in the underlying 
density field. The RIS then takes over, and again all models follow a
rather similar locus of $\xi_{HI}$ parameters. We have seen above that
the variation between the  $r_{0}-\gamma$ lines is mostly due
to the variation of $\gamma$ with $x_{m}$ in the different models.

 Overall, the change in slope for the different
models can be qualitively explained  by the different length scales of
the RIS features that occur in each, even at the same stages of reionization
($x_{m}$ values). There is a competition between small and large scale 
features that sets the slope of the correlation function. The behaviour
of $r_{0}$ is more puzzling, and we shall return to these
questions in Section 6.2

\subsection{The growth factor of perturbations}

\begin{figure}
\centerline{
\psfig{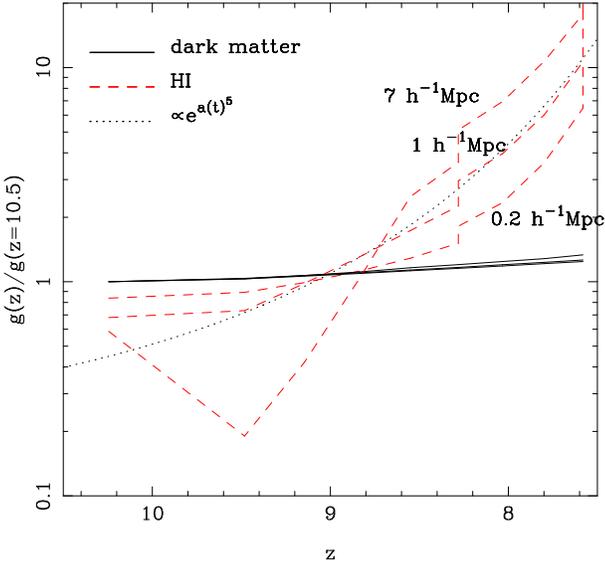}
}
\caption{The growth factor of perturbations in the fiducial model.
We show results for the gas density field as solid lines and the HI
density field as dashed lines, with a different
line for 3 different comoving scales. In each case we normalize by the
amplitude of fluctuations at redshift $z=10.5$. The dotted line
is a curve with $g \propto e^{a(t)^{5}}$.
\label{gfid}
}
\end{figure}

We have seen in previous sections that once reionization is visually 
progressing, the amplitude of fluctuations in the HI field grows extremely
rapidly. It is interesting to compare this quantitatively to the
growth expected of perturbations under gravitational instability. 
In the latter case, the amplitude of linear density perturbations
will grow at the same rate independent of scale, so that  
$\xi_{\rho}(r)$ will retain the same shape, but increase in amplitude
by an overall factor $[g(z)]^{2}$ which can be computed from first order 
perturbation theory (e.g., Peebles 1980). At the high redshifts relevant
here ($z \simgt 7$), the LCDM universe behaves in a manner
very close to an Einstein de Sitter model so that the linear growth factor
$g(z) \propto \frac{1}{1+z}$. From Figure \ref{xi_m}, we have seen that 
$r_{0}$ for the matter correlation function is between 0.1 and 0.5 $\hmpc$
for  the models over the range of redshifts when reionization occurs,
so that linear theory should be accurate over most of the range of scales
indicated by our resolution and boxsize tests (Section 5.1). We can 
also see from  Figure \ref{xi_m} that $\xi_{\rho}(r)$ does keep the same
shape, and increases in amplitude slowly, as expected.  This can also 
be inferred from Figure \ref{gfid}, where we plot the square root
of the ratio of the correlation function $\xi_{\rho}(r)$
 at redshift $z$ to that at 
redshift $z=10.5$, as a function of $z$. This is proportional to the
growth factor of pertubations between redshifts. There are three lines 
on the plot, corresponding to $r=0.2 \hmpc, r=1 \hmpc$ and $r=7 \hmpc$. The
lines are close together, indicating that the shape 
of $\xi_{\rho}(r)$ does not change dramatically.

In Figure \ref{gfid}, we plot the same quantity for $\xi_{HI}(r)$, for the
same $r$ values, for the fiducial model of reionization.
 In this case, we see both that growth of pertubations
is much more rapid, due to the RIS, than the linear density growth.
We draw a smooth curve corresponding to $g(z) \propto
e^{a(t)^5}$  alongside the simulation growth factor for $r= 1\hmpc$
in order to show roughly how extreme the growth in fluctuations
as a function of redshift is during the epoch of reionization. The
growth is faster than exponential over the short interval between
$z\sim 10$ and $z\sim 8$, it is roughly
an exponential function of a power law.

 We also
see that the different scales exhibit different growth rates, with the
large-scale fluctuations changing most rapidly. On $7 \hmpc$ scales,
the fluctuations are first suppressed, with the square root of $\xi_{HI}(r)$
decreasing by a factor $\sim 5$, before rapidly increasing after
$z=9.5$. This stronger behaviour relative to the smaller scales results
in the flattening of the correlation function. The roughly parallel
nature of the curves after $z \sim 8.5$ indicates that at the late stages
of reionization the flatter power law form of $\xi_{HI}(r)$ has been 
reached, and the amplitude grows similarily on all scales.

\begin{figure}
\centerline{
\psfig{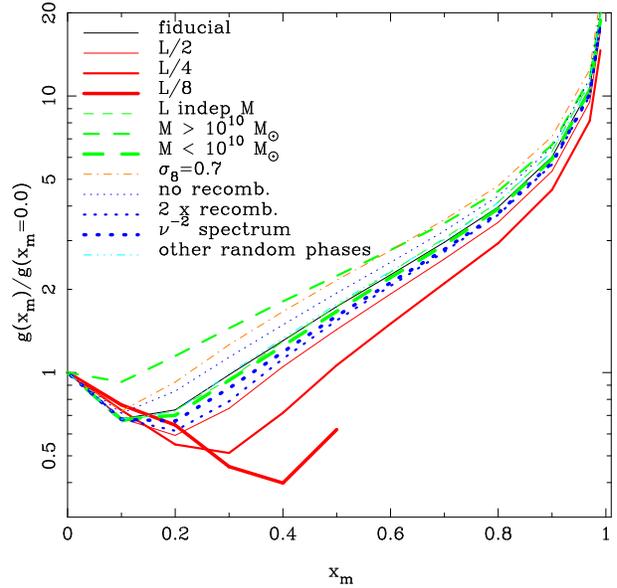}
}
\caption{The growth factor  of perturbations in the HI density
vs $x_{m}$ for all 12 models described in \S 2.3.
We show results for the 1 $\hmpc$ scale, and have normalized each
curve by the amplitude of perturbations measured at the last output time to 
be completely neutral.
\label{gm}
}
\end{figure}

If we now consider the 
growth of perturbations in the different models, a better way to compare them
is to look at the growth as a function of $x_{m}$. In Figure \ref{gm} we 
show this for the $r=1\hmpc$ scale. As the models become more ionized,
from left to right, the amplitude of fluctuations first dips and then rises
steeply. The model which dips the least on these scales is
the $M > 10^{10} \msun$ model, as we expect from looking at Figure \ref{xi_m}.
The main growth phase of fluctuations starts between $x_{m}=0.1$
$x_{m}=0.4$, depending on the model, with an exponential relationship
between $g$ and $x_{m}$. The curves appear to converge towards
the end of reionization, so that all models exhibit roughly the same amount of
growth, within $\sim 50 \%$ from the start of reionization until
$x_{m}=0.97$. The models which started off with less growth in $\xi_{HI}$ 
therefore have steeper dependence on $x_{m}$.
We find that $g\propto e^{2.8x_{m}}$ approximately holds over the
range $x_{m}=0.2-0.8$ for the fiducial model. 
The steepest curve has $g\propto e^{3.5x_{m}}$  (the L/4 model) and 
the shallowest has $g\propto e^{2.2x_{m}}$ (for the L $> 10^{10} \msun$ model).

If we look at the growth factor from the point of view of the parameters
varied in each model, we can see that decreasing the luminosity (from the
fiducial model through $L/2$, $L/4$ and $L/8$) monotonically changes
how abrupt the growth of HI fluctuations is during the bulk of the
reionization process. The model with least luminous sources ($L/8$) exhibits
fastest growth with respect to $x_m$, although of course with respect
to $z$, this is not necessarily the case. We have examined the growth
factor vs $z$ for the different models (not plotted) and find that 
the growth of fluctuations for all models has (at least for the
$1 \hmpc$ scale) a form roughly consistent wit the  $g(z) \propto
e^{a(t)^5}$  curve drawn on Figure \ref{gfid}. To zeroth order, the growth of
HI fluctuations as a function of $z$  does not seem to be dependent on 
the source physics.

\subsection{Babul and White model fit}

We have seen that a simple power law fit to the HI
 correlation function works well on scales for which the simulation has
sufficient boxsize and spatial resolution. As reionization proceeds, the
power law becomes shallower, moving away from the slope of the underlying
matter correlation function. While this paper's focus is on numerical
modelling and phenomenology of HI clustering rather than analytic modelling,
may neverthless be instructive to consider a different fitting function,
for $\xi_{HI}(r)$, derived from very simple model. Many analytic models
of bubble growth during reionization, of varying complexity 
(e.g., using the excursion set formalism, Furlanetto \etal 2004,
and perturbation theory, Zhang \etal 2007.)
 have been proposed, and most recently, tests
of the model of Furlanetto, Zhan, and Hernquist have been carried out by
Zahn \etal (2007) using
RT simulations.  Here we consider the model of Babul \& White (1991)
originally proposed
by those authors as a simple description of the way galaxy clustering
could be modulated by ``spheres of avoidance'' around quasars.

\begin{figure}
\centerline{
\psfig{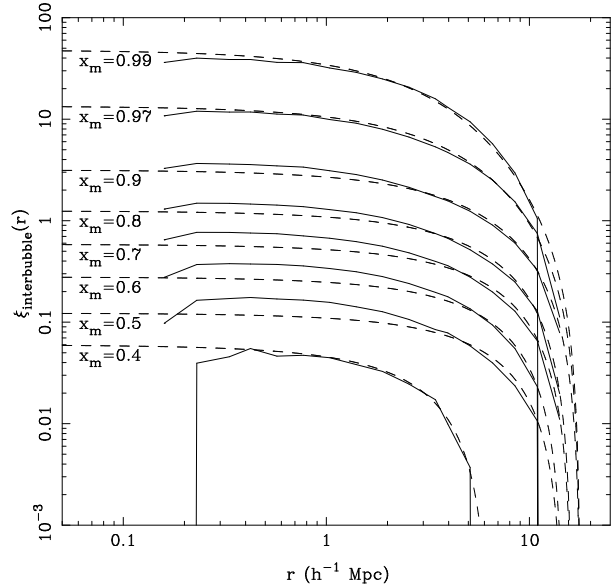}
}

\caption{
A fit of the functional form taken from Babul \& White (1991)
to the autocorrelation function of interbubble material 
(see Equation \ref{modu}).
We show results for the fiducial simulation, for 8 output times
with different values of the mean mass weighted ionized H density,
$x_{m}$. The fits were carried out to 
points with $r>0.3 \hmpc$.
\label{babul1}
}
\end{figure}

The model makes the simplifying assumption that the sources are
Poisson distributed and that they give rise to spheres of avoidance of 
a fixed physical size. The two point correlation function of
material distribution uniformly in the interbubble regions is then 
(Babul \& White 1991):

\begin{eqnarray}
1+\xi_{\rm interbubble}(r)=\nonumber \\
\exp\left( \frac{f_{b}}{2}\left
[\left(\frac{r}{2R_{b}}\right)^{3}-3\left(\frac{r}{2R_{b}}\right)
+2\right]\right)~~~~~ r  \leq 2R_{b} \nonumber \\
=1 ~~~~~ r> 2R_{b} 
\label{interbubble}
\end{eqnarray}

Here $f_{b}=4\pi n_{b}R_{b}^{3}/3$ is the nominal filling factor of spherical
bubbles of radius $r_{b}$ and mean number density $n_{b}$. If we
furthermore assume that the distribution of sources is independent
of the density distribution, modulation of the HI by bubbles leads to an HI
correlation function given by:

\begin{equation}
1+\xi_{HI}(r)=[1+\xi_{\rm interbubble}][1+\xi_{\rho}]
\label{modu}
\end{equation}

Of course, the distribution of sources {\it is} closely related to
the density distribution, something that leads to the antibiasing
seen in the early stages of reionization (see Figure \ref{bfid}).
 At later times, however,
this correlation may become less important in governing $\xi_{HI}$,
and the simple interbubble model may be useful. 

Rather than computing
values of the $f_{b}$ and $R_{b}$ parameters from theory, we use equation
\ref{interbubble} to fit $\xi_{\rm interbubble}$ measured from the
simulations, allowing the values of these parameters to vary. Before
fitting, we first compute $\xi_{\rm interbubble}$ using the
measured value of $\xi_{\rho}(r)$ in the simulation and
Equation \ref{modu}. $\xi_{\rm interbubble}$  for the fiducial
model for various values of $x_{m}$ is shown in Figure \ref{babul1}. For
$x_{m} < 0.4$,  $\xi_{\rm interbubble}$ is negative, an unphysical result
which occurs because of the correlation between sources and $\rho$.
 As reionization proceeds, $\xi_{\rm interbubble}$ becomes
dramatically larger in amplitude, and the right hand cut off
moves to larger scales. The correlation function is close to flat on smaller
scales than the break.

We can interpret this behaviour approximately in the context of the
Babul and White model fit, which is also shown in Figure \ref{babul1}.
We have fit the curves in the same manner as was carried out for the
power law fits in Section 5.4 (assuming Poisson errors).
It is evident that the simple model can reproduce the rough shape of
$\xi_{\rm interbubble}$, with agreement becoming best towards the end
stages of reionization. It is the modulation of the density correlation 
function by this form for the RIS $\xi$ results in the flattening
of $\xi_{HI}(r)$.

It would be interesting if the best fit value of $R_{b}$ for a given
$\xi_{\rm interbubble}$ could be used as a measure of the bubble size.
There are two problems with this, however. The first is that in the early
stages of reionization when bubbles are better defined, they are strongly
correlated with sources, so that the fit doesn't work. The second problem
is that at late times, the simulation box size we have used becomes
comparable to the bubble size and the results will presumably not be very 
reliable. There is a range of reionization stages for which this
might be useful though. For the fits in figure \ref{babul1} 
the $R_{b}$ values did vary 
monotonically with $x_m$, with $R_{b}=3.2 \hmpc$ for $x_{m}=0.4$, 
$R_{b}=7.4 \hmpc$ for $x_{m}=0.5$ and  $R_{b}=8.1 \hmpc$ for $x_{m}=0.7$. 
Above  $x_{m}=0.8$, $R_{b}$ saturated at a value of $9.0 \hmpc$,
possibly indicating the limitations of finite boxsize.

The other parameter, the filling factor $f_{b}$ can also be examined,
and could also be a useful diagnostic. In this case the value steadily
increases, with $f_{b}=0.06, 0.12, 0.25, 0.46, 0.81, 1.4, 2.7, 3.9$
for $x_{m}=0.4, 0.5, 0.6, 0.7, 0.8, 0.9, 0.97, 0.99$. We note from the
definition of $f_{b}$ above that it is a nominal filling factor, and as
a result $f_{b}$ will become greated than unity as bubbles overlap in the
late stages of reionization.

\begin{figure}
\centerline{
\psfig{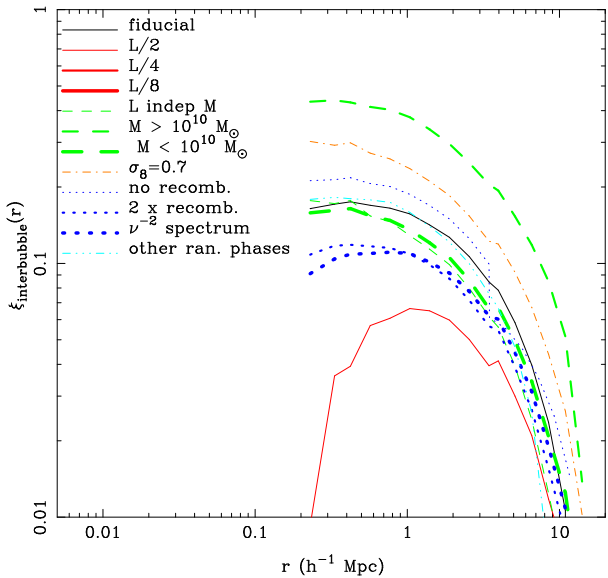}
}
\caption{
The autocorrelation function of interbubble material 
(see Equation \ref{modu}).
We show results for the 12 simulations described
in \S2.3, for the output times
when the mean mass weighted ionized H density,
$x_{m}$=0.5. 
\label{babul2}
}
\end{figure}

The difference between 
$\xi_{\rm interbubble}$ for the different models (all with $x_m=0.5$)
is shown in Figure \ref{babul2}. We can see that the peak amplitude of
the curves varies by a factor of $\sim 6$ between models, but that they
all display the same shallow curves behaviour with a cutoff. As we would
expect, the model with the largest obvious bubbles in Figure \ref{slice_m},
$M>10^{10}\msun$ has the largest amplitude and largest cutoff scale. A few
of the models (for example L/4 and L/8)  have a negative 
$\xi_{\rm interbubble}$ and are not plotted. For clarity, we do not 
show the model fits, but we note that the $M>10^{10}$
has the largest value of $f_{b}$, 0.27, and the L/2 model has the
smallest value, 0.06. These values for the bubble filling factor
appear by eye to be at least indicative of the trends seen in 
Figure \ref{slice_m}.

\section{Summary and discussion}

\subsection{Summary}

In this paper we have explored the radiation-induced 
structures (RIS) produced during the epoch of reionization and compared
them to the results of gravitational instability. Our conclusions
can be summarized as follows:\\

\noindent(1) For most of
the models we have tested, we find that the mean ionized
fraction of hydrogen increases exponentially,
$x_{m}=e^{-(z-z_{i})}$ (where  $z_{i}$ is the redshift of full
ionization)
as reionization proceeds.  For models adjusted to have emitted
the same number of source ionizing photons by $z=6$, there is still
quite a wide spread in the redshift of reionization, with models 
reaching $x_{m}=0.5$ from $z=9.0$ to $z=6.4$ (the last one to reionize
is the low $\sigma_{8}$ model.)\\
\noindent(2)  At a fixed $x_{m}$, the morphology of the RIS is most 
strongly affected by the lower cutoff in source luminosity, which 
changes the size of bubbles. The mass to light ratio of sources also has a
substantial effect, but the recombination rate and the amplitude
of mass fluctuations, $\sigma$ only minimally change the appearance
of the HI density field.\\
\noindent(3) The HI correlation function, $\xi_{HI}$ exhibits a generic
behaviour for all models tested. $\xi_{HI}$ initially becomes
linearly antibiased with respect to the matter $\xi_{\rho}$, as the 
high density HI around sources is ionized. The linear bias factor 
reaches a minimum of $b \sim 0.5$ when $x_{m} \sim 0.1-0.2$. The amplitude
of   $\xi_{HI}$  then increases rapidly, and  $\xi_{HI}$  keeps a scale
invariant shape, but the power law index flattens to an asymptotic
value of $\gamma \simeq -0.5$.\\
\noindent(4) We find that $r_{0}$,
 the correlation length of  $\xi_{HI}$ 
has the essentially the same functional relationship with $x_{m}$ in 
all but one of the models we have tested. How the power-law index  $\gamma$
varies with $x_{m}$
on the other hand  depends much more widely on the different source
and physics prescriptions adopted.\\
\noindent(5) The growth factor of HI perturbations
is seen to change much more rapidly than that of gravitionally evolving matter
perturbations over a redshift range $\Delta z \sim 2-3$ during which
the bulk of reionization occurs. 
We find pertubations on a scale of $1 \hmpc$ to be evolving
$\propto e^{a(t)^{5}}$ compared to $\propto a(t)$ for gravitational
growth. This is valid for all models tested, so that the source 
physics does not appear to affect relation.\\
\noindent(6) During the late stages of reionization, 
the shape evolution of  $\xi_{HI}$ can be approximately reproduced by a
simple model due to Babul \& White (1991) in which ionizing sources
are uncorrelated with the density field
and produce spherical bubbles. Fitting the parameters
of this model to $\xi_{HI}$ therefore form a method for inferring 
simple morphological characteristics from measurements of $\xi_{HI}$.\\

\subsection{Discussion}

In this paper, we have largely avoided discussing directly observational
probes of the RIS and the reionization epoch. This said, many of our findings
can be related closely to the possible results 
of a survey of 21cm brightness, as a function of 
angular position and wavelength. We have concentrated mainly on the
correlation function of the HI, and so in the most likely
scenario in which there have been enough early X-ray sources to heat
up the primordial gas just before reionization occurs, its temperature
is higher than the CMB and we are in the emission regime. In this
case, the 21cm brightness temperature is  independent of the spin
temperature. The brightness temperature $T \propto x_{\rm HI} (1+\delta)$,
where  $\delta$ is the overdensity of hydrogen
and $x_{\rm HI}$ the neutral fraction (see e.g., Madau 1997,
Di Matteo \etal 2004,
Barkana 2007.)

Looking at the autocorrelation function of the HI, the power law behaviour
we see is striking. For structure growing through gravitational instablity,
the initial fluctuations are scale-invariant over a large range
and they evolve in a scale free manner (Peebles 1974, Davis \& Peebles
1977.)  In the case of RIS, the onset of significant clustering is
extremely rapid, and it might be expected that this would lead to
features in the correlation function related to for example the Stromgren
radius of sources dominant at that time. On the other hand,
cosmic variance would lead to a substantial scatter in the scale length
of features from place to place. Our fiducial simulation volume, at $40$
comoving $\hmpc$
side length also limits our ability to capture the late stages
of reionization. Wyithe and Loeb (2004) have predicted a comoving 
radius of bubbles $\sim 40 \hmpc$ at the end of the overlap stage.
We have seen in \S 5.6 that the simple Babul and White model 
fit does capture the 
shape of $\xi(r)$ reasonably well even though there is a distinct bubble
scale in the model. Scale invariance therefore appears relatively easy
to achieve and detection of departures from it probably requires a 
wide range of scales to be available.

Furlanetto \etal (2006) have used the analytic model of Furlanetto \etal 
(2004) 
to describe how the HII bubble size is related to both the bias of galaxies
and the underlying matter power spectrum. The success of such analytic
models in comparisons to simulations (e.g., Zahn \etal 2007) has opened
up the way for their use in analyzing future 21cm observations.  Our
related work, attempting to directly 
simulate a relatively wide range of models has
found that the autocorrelation function of 21cm emission can be used to infer
the broad signatures of RIS compared to gravitational structure  (e.g.,
non-monotonic
growth, flatter asymptotic slope $\gamma \simeq -0.5$) which will help in
analysis of the first observations. 

 Among the other analytic models
which have been developed, the pertubation theory approach of
Zhang \etal (2007) is different from many in that it does not make a step 
function bubble approximation to the HI distribution. As our simulation
approach includes recombinations and the contribution of residual
HI in the ionized regions, future comparisons will be beneficial. For example,
Zhang \etal (2007) compute the rapid rise in the bias of HI clustering
as a function of redshift, finding some qualitatively similar
results to ours, although they find a scale-independent bias on 
large scales.

One  aspect which we have not covered is the effect of
redshift distortions on the RIS. We expect the autocorrelation 
function to be affected by the usual line of sight amplification 
(Kaiser 1987) on large scales, and small scale suppression
from the velocity dispersion 
(e.g. Peebles 1980) on small scales. The latter effect in particular
is likely to be strongly modified by the fact that most of the HI around 
bright sources and hence in dense regions is ionized early on. These effects
will likely be important in the use of 21cm emission maps to carry
out tests of cosmic geometry (see e.g., Nusser 2005.) It would be simple enough
in future work to look at the simulations in redshift space, such as would be 
seen with observational data.

In carrying out our numerical experiments, we have simulated a range of
models, which we expect to have many of the features
likely in most scenarios for the reionization of the Universe. It was not 
possible to be completely general, however, and it is certainly possible
to imagine other interesting models and tests of the physics that could
be included. For example, McQuinn \etal (2007) ran as one of their
many models one in which all cells in the simulation were set to the mean
density. This had the effect of changing the amount of structure in the
ionization fronts, and of course drastically reducing the recombination
rate.

With our postprocessing RT carried out on the
hydrogen distribution only, by keeping track
of the ionization state but not predicting the temperature we
necessarily simplified much of the physics involved in reionization. As our
intention was to capture the broad differences between RIS and gravitational
structure, we do not regard this as important. 
However in future work extending our simulation approach so that it is
directly relevant to upcoming observational data, we plan to capture more
detail. For example, the code {\small SPHRAY} (Altay \etal 20007) 
 which represents an extension of the
present method includes different ionized species and explicit temperature
evolution. Other work such as Ciardi \etal (2006) and McQuinn \etal (2007)
specifically include the effect of minihalos, sources below the resolution
scale of the simulation. The latter group finds that they have a strong effect
on the growth of large bubbles in the late stages of reionization.
In the present work, we have seen from our resolution studies that the
autocorrelation function at least is not sensitive to increases
in mass resolution and small scale structure, at least to the accuracy
and range of models that we have considered. In principle, the high spatial 
resolution of our gridless ray tracing approach could allow the effects
of Lyman-limit systems to be modelled, which become dominant in limiting
the photon mean free path when the universe is mostly ionized (e.g.,
Miralda-Escud\'{e} \etal 2000.)

Future work on 
realistic models should also include radiative cooling in the formation
of sources and modelling of 
specific sources and spectra.
Quasars and miniquasars have been investigated by including 
the formation and growth of black holes together with a model for 
feedback directly
into cosmological SPH simulations (Di Matteo \etal 2007, Sijacki \etal 2007,
Pelupessy \etal 2007). Using such models
as sources in RT calculations would enable us to investigate how the
RIS caused by harder sources is different to softer sources, in ways which
are not constrained by the simple association of source and halo
mass as has been carried out here.

\section*{Acknowledgments}
 This work was supported by the 
NASA Astrophysics Theory Program, contract NNG 06-GH88G
and NSF grant AST-0507665.

\bibliographystyle{mn}	

\end{document}